\documentclass[prd,preprintnumbers,amsmath,amssymb,nofootinbib]{revtex4}
\usepackage{graphics,color,subfigure}
\usepackage{epsfig}
\usepackage{dcolumn}% Align table columns on decimal point
\usepackage{bm}% bold math
\usepackage{longtable}
\usepackage{multirow}
\usepackage{diagbox}
\usepackage[colorlinks=true,
urlcolor=blue,
linkcolor=blue,
citecolor=blue,
bookmarks=true,
bookmarksopen=true,
bookmarksnumbered=true]{hyperref}

\begin{document}
	\title{Doubly-charmed pentaquark states in a mass splitting model}
	\author{Shi-Yuan Li$^1$}
	\author{Yan-Rui Liu$^1$}%\email{yrliu@sdu.edu.cn}
	\author{Cheng-Rui Shu$^1$}%\email{202212081@mail.sdu.edu.cn}
	\author{Zong-Guo Si$^1$}
	\affiliation{$^1$School of Physics, Shandong University, Jinan, Shandong 250100, China	}
	\date{\today}
	
\begin{abstract}
Concentrating on the mass differences relative to $P_{\psi}^{N}(4312)^+$, we systematically investigate the spectra of doubly-charmed pentaquark states in the compact $ccqq\bar{q}$ ($q=u, d, s$) configuration. The assumption that the observed $P_{\psi}^{N}(4312)^+$ is a compact hidden-charm pentaquark with $I(J^P)=\frac12(\frac32^-)$ is adopted. We also study the properties of strong decays within a simple rearrangement scheme. The results indicate that the $I(J^P)=\frac12(\frac12^-)$ $ccnn\bar{n}$ with $I_{nn}=0$ where $n$ denotes $u$ or $d$ quark, $I(J^P)=0(\frac12^-)$ $ccnn\bar{s}$, and $I(J^P)=0(\frac12^-)$ $ccns\bar{n}$ ground states should be stable. 
%The predicted properties can be tested in future experiments.
\end{abstract}
\maketitle

%==================================================
\section{Introduction}\label{secI}
%==================================================
	
After the discovery of the charmonium-like state $\text{X}(3872)$ \cite{Belle:2003nnu}, a series of hidden- and open-heavy exotic mesons have been identified as four-quark states over the last two decades \cite{BaBar:2005hhc,BaBar:2006ait,Belle:2007hrb,Belle:2011aa,Belle:2013yex,BESIII:2013ris,CDF:2009jgo,Belle:2009rkh,CDF:2011pep,LHCb:2016axx,Olsen:2017bmm,LHCb:2021uow,LHCb:2022aki,LHCb:2020bls,LHCb:2020pxc,LHCb:2022sfr,LHCb:2022lzp,LHCb:2021auc,LHCb:2021vvq,LHCb:2020bwg,CMS:2023owd,ATLAS:2023bft}. In addition to mesons, heavy flavor pentaquark-like baryon states have also garnered significant attentions. In 2015, two possible hidden-charm pentaquark states $P_{\psi}^N(4380)^+$ and $P_{\psi}^N(4450)^+$ were observed in the decay $\Lambda_b^0\rightarrow J/\psi pK^-$ by the LHCb Collaboration \cite{LHCb:2015yax}. Subsequently, in 2019, a new pentaquark candidate  $P_{\psi}^N(4312)^+$ was reported by the LHCb in the same decay mode, while the $P_\psi^N(4450)^+$ was resolved into two states, $P_\psi^N(4440)^+$ and $P_\psi^N(4457)^+$ \cite{LHCb:2019kea}. Recently, the collaboration announced the observation of another $J/\psi p$ resonance $P_\psi^N(4337)^+$, as well as two $J/\psi\Lambda$ resonances $P_{\psi s}^{\Lambda}(4459)^0$ and $P_{\psi s}^{\Lambda}(4338)^0$, in the decay channels $B_s^0\rightarrow J/\psi p\bar{p}$, $\Xi_b^-\rightarrow J/\psi \Lambda K^-$, and $B^-\rightarrow J/\psi\Lambda K^-$, respectively \cite{LHCb:2021chn,LHCb:2020jpq,LHCb:2022ogu}. In Ref. \cite{Belle:2025pey}, the Belle and BelleII Collaborations reported the evidence of $P_{\psi s}^{\Lambda}(4459)^0$ in $\Upsilon(1S,2S)$ inclusive decays. These experimental results have sparked heated discussions on their possible internal structures and physical properties. %Theoretical studies investigating their structures and properties have proliferated in response to these .
Various configurations are proposed to elucidate the nature of these unconventional hadrons, which include the compact multiquark picture, the hadron-hadron molecule picture, and the hybrid structure, etc. Detailed discussions on the theoretical studies can be found in relevant review articles \cite{Brambilla:2019esw,Chen:2016qju,Huang:2023jec,Liu:2019zoy,Lebed:2016hpi,Guo:2017jvc,JPAC:2021rxu,Ali:2017jda,Guo:2019twa,Meng:2022ozq,Chen:2022asf,Liu:2024uxn}.

The intriguing experimental findings also triggered discussions on possible doubly-charmed pentaquark states \cite{Zhou:2018bkn,Wang:2018lhz,Park:2018oib,Giannuzzi:2019esi,Xing:2021yid,Andreev:2022qdu,Ozdem:2022vip,Park:2023ygm,Yang:2024okq,Guo:2017vcf,Chen:2017vai,Dias:2018qhp,Yan:2018zdt,Shimizu:2017xrg,Yang:2020twg,Chen:2021kad,Zhou:2022gra,Wang:2022aga,Shen:2022zvd,Yalikun:2023waw,Wang:2023mdj,Wang:2023eng,Liu:2023clr,Duan:2024uuf,Wang:2024brl}. Replacing the light quark in the conventional baryon $\Xi_{cc}$ with two light antiquarks, one gets a doubly-charmed tetraquark $T_{cc}$ which was predicted in various theoretical models long time ago \cite{Ballot:1983iv,Zouzou:1986qh}. Replacing one light antiquark in the $T_{cc}$ with two quarks further, or putting a $q\bar{q}$ pair inside the $\Xi_{cc}$, one gets a potential doubly-charmed pentaquark state.  To date, no such experimental candidate has been announced. However, the existence of such states cannot be excluded. Insights may be drawn from the search for $\Xi_{cc}$ and $T_{cc}$. Two decades ago, the SELEX Collaboration claimed the observation of the quark-model-predicted doubly-charmed baryon $\Xi_{cc}^+$ in the $\Lambda_c^+K^-\pi^+$ three-body decay channel \cite{SELEX:2002wqn}, but other collaborations did not confirm its existence. In 2017, its partner state $\Xi_{cc}^{++}$ was observed by the LHCb \cite{LHCb:2017iph} in the $\Lambda_c^+K^-\pi^+\pi^+$ channel with a $\sim$100 MeV higher mass than the SELEX $\Xi_{cc}^+$. The LHCb also observed $\Xi_{cc}^{++}$ in the channel $\Xi_{c}^+\pi^+$ \cite{LHCb:2018pcs}. Later in 2022, the LHCb succeeded in finding a doubly-charmed four-quark state, $T_{cc}^+(3875)$, in the $D^0D^0\pi^+$ mass distribution just below the $D^{*+}D^0$ threshold \cite{LHCb:2021auc,LHCb:2021vvq}.  The observations of $\Xi_{cc}^{++}$ and $T_{cc}^+(3875)$ lend credibility to find the doubly-charmed pentaquarks finally, if they should exist from a theoretical point of view.  

In the literature, a number of discussions on $QQqq\bar{q}$ ($Q=c,b$; $q=u,d,s$) with respect to their spectra, decays, electromagnetic properties, and so on in configurations such as compact pentaquark \cite{Zhou:2018bkn,Wang:2018lhz,Park:2018oib,Giannuzzi:2019esi,Xing:2021yid,Andreev:2022qdu,Ozdem:2022vip,Park:2023ygm,Yang:2024okq} and hadronic molecule \cite{Guo:2017vcf,Chen:2017vai,Dias:2018qhp,Yan:2018zdt,Shimizu:2017xrg,Yang:2020twg,Chen:2021kad,Zhou:2022gra,Wang:2022aga,Shen:2022zvd,Yalikun:2023waw,Wang:2023mdj,Wang:2023eng,Liu:2023clr,Duan:2024uuf,Wang:2024brl,Xu:2010fc,Sheng:2024hkf,Yang:2024okq} can be found. The electromagnetic properties were studied in Refs. \cite{Zhou:2022gra,Wang:2023ael,Ozdem:2024yel,Ozdem:2022vip}. In Refs. \cite{Zhou:2018bkn,Park:2018oib,Park:2023ygm}, the authors explored spectra in the compact pentaquark picture utilizing color-magnetic interaction (CMI) models, while QCD sum rules were adopted for the study of compact states in Ref. \cite{Wang:2018lhz,Yang:2024okq}. A quark potential model was employed in a diquark-diquark-antiquark configration to estimate the masses in Ref. \cite{Giannuzzi:2019esi}. The authors of Ref. \cite{Xing:2021yid} obtained spectra of $ccqq\bar{q}$ states in a triquark-diquark configuration and calculated decay widths using the operator product expansion technique. Compared with the compact configuration, the baryon-meson molecule is a more popular one in exploring doubly-heavy pentaquark-like states. Such studies were performed with various methods, like one-boson-exchange potential model \cite{Shimizu:2017xrg,Chen:2017vai,Dias:2018qhp,Chen:2021kad,Dong:2021bvy,Shen:2022zvd,Wang:2023aob,Yalikun:2023waw,Xu:2010fc}, chiral quark model \cite{Yang:2020twg}, chiral effective field theory \cite{Xu:2010fc,Guo:2017vcf,Yan:2018zdt,Chen:2021htr}, QCD sum rules \cite{Duan:2024uuf,Wang:2024brl,Yang:2024okq}, local hidden gauge approach \cite{Wang:2022aga,Wang:2023mdj}, resonating group method based on the QDCSM framework \cite{Liu:2023clr}, near-threshold interaction with the light $qq$ and $q\bar{q}$ correlations \cite{Wang:2023eng}, etc. %These theoretical articles utilized various approaches to investigate doubly charmed pentaquark, enhancing our understanding of exotic hadrons.
	
To understand basic features about their spectra and decay widths, we here investigate the $S$-wave compact $ccqq\bar{q}$ states in a modified CMI model. Previously in Ref. \cite{Zhou:2018bkn}, the pentaquark masses were estimated with two formulae in the CMI framework, the original one using effective quark masses and the modified one using a meson-baryon threshold. Their applications to the hidden-charm pentaquarks indicate that the latter formula gives more reasonable results in understanding the LHCb $P_\psi^N$ states in the compact configuration \cite{Wu:2017weo,Cheng:2019obk}. The main reason should be that the appropriate inclusion of quark interactions in the CMI model investigation of pentaquark properties is reflected in the selection of scale parameters, quark masses or meson-baryon thresholds. A problem arises, however, in selecting an appropriate meson-baryon threshold with the latter formula because more than one possibilities exist. To alleviate this problem, in Ref. \cite{Li:2023aui}, a third mass formula was introduced by choosing the scale parameter as the mass of $P_\psi^N(4312)$ which is assumed to be a compact $uudc\bar{c}$ pentaquark. Combining the resulting mass and decay information, one may understand the properties of LHCb $P_\psi^N$ and $P_{\psi s}^\Lambda$ states in the compact configuration. The consistency interpretation requires that one has to assign the spin-parity of $P_\psi^N(4312)$ to be $J^P=\frac32^-$, which is different from the molecule interpretation $J^P=\frac12^-$. In the present study, we consider the doubly-charmed case using the third type formula. Additionally, we investigate strong decay widths in a simple rearrangement scheme, which was not discussed in Ref. \cite{Zhou:2018bkn}.

This article is arranged as follows. The formalism including three mass formulae, spin-color base vectors, and rearrangement decays, is presented in Sec. \ref{secII}. The parameters for calculation and the numerical results are listed in Sec. \ref{secIII}. Finally, some discussions and a brief summary are provided in the last section.
	
%==================================================
\section{Formalism}\label{secII}
%==================================================

\subsection{Mass formulae}\label{secIIA}
	
The simple CMI model used to study the $S$-wave $ccqq\bar{q}$ states has the Hamiltonian
\begin{eqnarray}\label{hamiltonian}
H=\sum_i m_i+H_{CMI}=\sum_i m_i-\sum_{i<j}C_{ij}\vec{\lambda}_i\cdot\vec{\lambda}_j\vec{\sigma}_i\cdot\vec{\sigma}_j.
\end{eqnarray}
Here, the subscript $i$ represents the $i$th constituent quark. The effective quark mass, denoted as $m_i$, contains contributions from various terms such as quark kinetic energy, color Coulomb interaction, and color confinement. The parameter $C_{ij}$ is the effective coupling constant characterizing the interaction strength between the $i$th and $j$th quark components. One extracts these parameters from the measured masses of conventional hadrons. $\lambda_i$ and $\sigma_i$ denote the $SU(3)$ Gell-Mann matrices and $SU(2)$ Pauli matrices for the $i$th quark, respectively. For the antiquark, $\lambda_i$ should be replaced by $-\lambda_i^*$. After one establishes the spin-color base vectors, the CMI matrix $\langle H_{CMI}\rangle$ and then its eigenvalues $E_{CMI}$'s and eigenvectors can be determined. It is easy to write down the mass formula corresponding to Eq. \eqref{hamiltonian}
\begin{eqnarray}\label{mass1}
M=\sum_im_i +E_{CMI}.
\end{eqnarray}
Because hadrons have different internal quark interactions, applying the extracted $m_i$ and $C_{ij}$ to multiquark systems in the CMI model may induce non-negligible uncertainties on the mass spectra. Previous numerical analyses \cite{Liu:2019zoy,Wu:2018xdi,Cheng:2020nho,Li:2023wxm,Li:2023wug} have shown that Eq. \eqref{mass1} usually yields overestimated values which are mainly due to the effective quark masses. We will regard the results estimated with this formula as theoretical upper limits for  pentaquark masses. 
	
More reasonable spectra may be obtained if one replaces the scale determined by effective quark masses with a reference scale. In the pentaquark case, a possible choice of reference scale is the threshold of a meson-baryon state which has the same quark content as the considered pentaquark. Now, the mass formula reads
\begin{eqnarray}\label{mass2}
M=[M_{ref}-(E_{CMI})_{ref}]+E_{CMI},
\end{eqnarray}
where $M_{ref}$ and $(E_{CMI})_{ref}$ are the employed meson-baryon threshold and related CMI eigenvalue, respectively. The scale replacement indicates that the needed attraction using Eq. \eqref{mass1} is effectively incorporated by the quark interactions inside the reference state. However, in the general case, two or more different thresholds may be adopted for a considered system and it is a problem to determine which one is more appropriate. Besides, previous studies in the tetraquark case \cite{Wu:2018xdi,Cheng:2020nho,Li:2023wxm,Li:2023wug} have shown that the estimated masses with meson-meson thresholds may be lower than the experimentally measured values and one can regard such results as theoretical lower limits. We will discuss whether it is also the case for doubly-charmed pentaquark states.

It is generally considered that hadron-hadron molecules and compact multiquarks have different quark interactions. In the situation that different compact multiquarks have better similarities in inner structures than different configurations (molecule and multiquark), one anticipates reduced uncertainties in mass predictions by introducing a reference multiquark state. It is better to use just one scale parameter for all the multiquark systems. Then a third mass formula can be obtained. In the tetraquark case, we chose to utilize the mass of the reference $\chi_{c1}(4140)$ by treating it as a compact $cs\bar{c}\bar{s}$ state with $J^{PC}=1^{++}$ \cite{Wu:2018xdi}. Subsequent studies with this assumption \cite{Cheng:2020nho,Li:2023wxm,Li:2023wug} indicated that the properties of exotic mesons like $X(3960)$ and $T_{cc}(3875)$ may be understood in the compact tetraquark picture. In the present case, we similarly adopt a reference pentaquark state. When studying the hidden-charm $P_\psi^N$ and $P_{\psi s}^\Lambda$ states in the compact configuration \cite{Cheng:2019obk,Li:2023aui}, we identified their possible structures by analyzing pentaquark spectra and decay widths. To give a consistent interpretation, one has to assign the quantum numbers of $P_\psi^N(4312)^+$ to be $I(J^P)=\frac12(\frac32^-)$ while they are $I(J^P)=\frac12(\frac12^-)$ in the molecule configuration. For studies of compact pentaquarks, we select $P_{\psi}^N(4312)^+$ with $I(J^P)=\frac12(\frac32^-)$ as the reference state and the new mass formula is
\begin{eqnarray}\label{mass3}
M&=&[M_{P_\psi^N(4312)}-(E_{CMI})_{P_\psi^N(4312)}]+E_{CMI}+\sum_{i,j=u,d,s,c,b}n_{ij}\Delta_{ij}\\\nonumber
&=&\tilde{m}_{penta}+E_{CMI}+\sum_{i,j=u,d,s,c,b}n_{ij}\Delta_{ij}.
\end{eqnarray} 
Here, $M_{P_\psi^N(4312)}$ and $(E_{CMI})_{P_\psi^N(4312)}$ are the measured mass and calculated CMI eigenvalue of $P_{\psi}^N(4312)^+$, respectively, and $\tilde{m}_{penta}\equiv M_{P_\psi^N(4312)}-(E_{CMI})_{P_\psi^N(4312)}$. The last term accounts for the scale discrepancy when the studied system possesses different quark contents from $P_{\psi}^N(4312)^+$. $\Delta_{ij}=m_i-m_j$ signifies the mass gap between two quarks and $n_{ij}$ is an integer. $\Delta_{ij}$'s will be treated as constant parameters and they may be determined from conventional hadrons. Explicitly, the formulae to be used in the following discussions are ($n=u,d$)
\begin{eqnarray}\label{mass3detail}
M_{ccnn\bar{n}}&=&\tilde{m}_{penta}+E_{CMI},\nonumber\\
M_{ccnn\bar{s}}&=&\tilde{m}_{penta}+E_{CMI}+\Delta_{sn},\nonumber\\
M_{ccns\bar{s}}&=&\tilde{m}_{penta}+E_{CMI}+2\Delta_{sn},\\
M_{ccss\bar{s}}&=&\tilde{m}_{penta}+E_{CMI}+3\Delta_{sn},\nonumber\\
M_{ccns\bar{n}}&=&M_{ccnn\bar{s}},\quad  M_{ccss\bar{n}}=M_{ccns\bar{s}}.\nonumber
\end{eqnarray}
Although two different systems may have the same mass formula, the resulting spectra are actually distinct. We will regard the masses calculated with Eq. \eqref{mass3detail} as the theoretical predictions in this article.

\subsection{Spin-color base vectors}\label{secIIB}

The calculations of $\langle H_{CMI}\rangle$, $E_{CMI}$, and rearrangement decay widths need the flavor-spin-color wave functions that are constrained by the Pauli principle. We do not consider the isospin breaking effects and the systems to be studied are $ccnn\bar{n}$, $ccnn\bar{s}$, $ccss\bar{n}$, $ccss\bar{s}$, $ccns\bar{n}$, and $ccns\bar{s}$.

In color space, one has three base vectors,
\begin{eqnarray}\label{color}
\phi_{AA}&=&((cc)_{\bar{3}}(qq)_{\bar{3}}\bar{q}_{\bar{3}})_1,\nonumber\\
\phi_{AS}&=&((cc)_{\bar{3}}(qq)_6\bar{q}_{\bar{3}})_1,\\
\phi_{SA}&=&((cc)_{6}(qq)_{\bar{3}}\bar{q}_{\bar{3}})_1,\nonumber		
\end{eqnarray}
where $A$ and $S$ mean anti-symmetry and symmetry, respectively, under the exchange of two identical quarks. The first (second) subscript of $\phi$ is for the heavy (light) diquark. The subscripts 1, 6, and $\bar{3}$ are the representations of color $SU(3)$. By using the Clebsch-Gordan coefficients \cite{deSwart:1963pdg,Kaeding:1995vq}, the color wave functions can be explicitly expressed. The detailed bases utilized in our work are aligned with those in Ref. \cite{Zhou:2018bkn} and we do not list them here. In spin space, the categorized base vectors read
\begin{eqnarray}
\chi_{SS}&:&\left\{\begin{array}{ccc}\chi_1&=&[((cc)^1(qq)^1)^2\bar{q}]^{\frac52},\\ 
\chi_2&=&[((cc)^1(qq)^1)^2\bar{q}]^{\frac32},\\
\chi_3&=&[((cc)^1(qq)^1)^1\bar{q}]^{\frac32},\\
\chi_4&=&[((cc)^1(qq)^1)^0\bar{q}]^{\frac12},\\
\chi_5&=&[((cc)^1(qq)^1)^1\bar{q}]^{\frac12},\end{array}\right.\nonumber\\
\chi_{SA}&:&\left\{\begin{array}{ccc}\chi_6&=&[((cc)^1(qq)^0)^1\bar{q}]^{\frac32},\\ \chi_7&=&[((cc)^1(qq)^0)^1\bar{q}]^{\frac12},\end{array}\right.\\
\chi_{AS}&:&\left\{\begin{array}{ccc}\chi_8&=&[((cc)^0(qq)^1)^1\bar{q}]^{\frac32},\\ \chi_9&=&[((cc)^0(qq)^1)^1\bar{q}]^{\frac12},\end{array}\right.\nonumber\\
\chi_{AA}&:&\chi_{10}=[((cc)^0(qq)^0)^0\bar{q}]^{\frac12},\nonumber
\end{eqnarray}
where the meaning of $A$ and $S$ is the same as $\phi$ while the superscripts 0, 1, 2, $\frac12$, $\frac32$, and $\frac52$ represent spins. Because of the Pauli principle, the allowed spin-color base vectors for $ccss\bar{q}$ and $(ccnn\bar{q})^{I_{nn}=1}$ are $\chi_{SS}\phi_{AA}$, $\chi_{SA}\phi_{AS}$, and $\chi_{AS}\phi_{SA}$, those for $(ccnn\bar{q})^{I_{nn}=0}$ are $\chi_{SS}\phi_{AS}$, $\chi_{SA}\phi_{AA}$, and $\chi_{AA}\phi_{SA}$, and those for $ccns\bar{q}$ are $\chi_{SS}\phi_{AS}$, $\chi_{SS}\phi_{AA}$, $\chi_{SA}\phi_{AS}$, $\chi_{SA}\phi_{AA}$, $\chi_{AA}\phi_{SA}$, and $\chi_{AS}\phi_{SA}$. We present details for each type state in table \ref{basis}. Since these wave functions are aligned with those in Ref. \cite{Zhou:2018bkn}, the resulting CMI matrices are the same and we do not repeat them here.
\begin{table}[htbp]
\caption{The color-spin base vectors of $ccqq\bar{q}$ pentaquark states.}\label{basis}
\begin{tabular}{|c|ccc|}\hline
\diagbox{$J^P$}{flavor}&$(ccss\bar{q})$, $(ccnn\bar{q})^{I_{nn}=1}$&$(ccnn\bar{q})^{I_{nn}=0}$&$(ccns\bar{q})$\\\hline
$\frac52^-$&$\chi_1\phi_{AA}$&$\chi_1\phi_{AS}$&$\begin{array}{c}\chi_1\phi_{AS}\\\chi_1\phi_{AA}\end{array}$\\\hline
$\frac32^-$&$\begin{array}{c}\chi_2\phi_{AA}\\\chi_3\phi_{AA}\\\chi_6\phi_{AS}\\\chi_8\phi_{SA}\end{array}$&$\begin{array}{c}\chi_2\phi_{AS}\\\chi_3\phi_{AS}\\\chi_6\phi_{AA}\end{array}$&$\begin{array}{c}\chi_2\phi_{AS}\\\chi_2\phi_{AA}\\\chi_3\phi_{AS}\\\chi_3\phi_{AA}\\\chi_6\phi_{AS}\\\chi_6\phi_{AA}\\\chi_8\phi_{SA}\end{array}$\\\hline
$\frac12^-$&$\begin{array}{c}\chi_4\phi_{AA}\\\chi_5\phi_{AA}\\\chi_7\phi_{AS}\\\chi_9\phi_{SA}\end{array}$&$\begin{array}{c}\chi_4\phi_{AS}\\\chi_{10}\phi_{SA}\\\chi_5\phi_{AS}\\\chi_7\phi_{AA}\end{array}$&$\begin{array}{c}\chi_4\phi_{AS}\\\chi_4\phi_{AA}\\\chi_{10}\phi_{SA}\\\chi_5\phi_{AS}\\\chi_5\phi_{AA}\\\chi_7\phi_{AS}\\\chi_7\phi_{AA}\\\chi_9\phi_{SA}\end{array}$\\\hline
\end{tabular}
\end{table}

\subsection{Rearrangement decay}

In present study, we consider just the two-body $S$-wave strong decays of pentaquark states which do not involve quark pair creation. If the Hamiltonian $H_{decay}$ is known for rearrangement decays, the calculation of the amplitude $\mathcal{M}=\langle final|H_{decay}|initial\rangle$ is possible. Since it is not easy to derive $H_{decay}$ theoretically, for simplicity, here we just assume that $H_{decay}$ is a constant denoted as ${\cal C}$. Consequently, the amplitude can be expressed as $\mathcal{M}={\cal C}\langle final|initial\rangle$. We also make the assumption that the physical total width of a pentaquark is equal to the sum of two-body partial decay widths, i.e. $\Gamma_{sum}\approx\Gamma_{total}$. Thus one may extract the parameter ${\cal C}$ from experimental data. The eigenfunction of an initial state $|initial\rangle$ can be represented as $\sum_ix_i|\psi_i\rangle$, where $|\psi_i\rangle$ and $x_i$ are the base vectors shown in table \ref{basis} and the corresponding coefficients, respectively. The final meson-baryon state can also be expanded similarly, $|final\rangle=\sum_jy_j|\psi_j\rangle$. Then $\mathcal{M}={\cal C}\sum_ix_iy_i$ is obtained and one gets the partial decay width for the given channel using the formula
\begin{eqnarray}\label{decay}
\Gamma=|\mathcal{M}|^2\frac{|\vec{p}|}{8\pi M^2},
\end{eqnarray}
where $M$ is the mass of the initial pentaquark and $\vec{p}$ denotes the three-momentum of the final meson or baryon in the rest frame of the initial state. 
	
Among all possible final hadrons, mesons with isospin equal to 0, i.e., $\eta$, $\eta'$, $\omega$, and $\phi$, are special due to the mixing between the flavor singlet wavefunction $\psi_1=\frac{1}{\sqrt{3}}(\sqrt{2}n\bar{n}+s\bar{s})$ and the flavor octet  wavefunction $\psi_8=\frac{1}{\sqrt{3}}(n\bar{n}-\sqrt{2}s\bar{s})$. For $J^P=1^-$ states $\omega$ and $\phi$, their mixing is nearly ideal and one has $\omega=n\bar{n}$ and $\phi=s\bar{s}$. However, for the $0^-$ states $\eta$ and $\eta'$, the mixing should be taken into account \cite{Li:2023wxm}. Their flavor wavefunctions are
\begin{eqnarray}
\left\{\begin{array}{ccccc}\eta&=&\psi_8\cos\theta-\psi_1\sin\theta&=&(\frac{1}{\sqrt{3}}\cos\theta-\sqrt{\frac23}\sin\theta)n\bar{n}-(\sqrt{\frac23}\cos\theta+\frac{1}{\sqrt{3}}\sin\theta)s\bar{s},\\
\eta'&=&\psi_8\sin\theta+\psi_1\cos\theta&=&(\frac{1}{\sqrt{3}}\sin\theta+\sqrt{\frac23}\cos\theta)n\bar{n}+(\frac{1}{\sqrt{3}}\cos\theta-\sqrt{\frac23}\sin\theta)s\bar{s},\end{array}\right.
\end{eqnarray}
where $\theta=-11.3^{\circ}$ \cite{ParticleDataGroup:2024cfk} is the mixing angle.

%==================================================
\section{Numerical results}\label{secIII}
%==================================================
\subsection{Model parameters}

When estimating the masses of doubly-charmed pentaquark states with Eq. \eqref{mass3detail}, we need to know the values of coupling parameters $C_{ij}$'s, $\Delta_{sn}$, and $\tilde{m}_{penta}$. From the masses of conventional hadrons, one gets $C_{nn}=18.4\ \text{MeV}$, $C_{ns}=12.1\ \text{MeV}$, $C_{cn}=4.0\ \text{MeV}$, $C_{ss}=5.7\ \text{MeV}$, $C_{cs}=4.3\ \text{MeV}$, $C_{n\bar{n}}=29.9\ \text{MeV}$, $C_{n\bar{s}}=18.7\ \text{MeV}$, $C_{c\bar{n}}=6.6\ \text{MeV}$, $C_{c\bar{s}}=6.7\ \text{MeV}$, and $C_{c\bar{c}}=5.3\ \text{MeV}$ with the extraction procedure shown in \cite{Wu:2018xdi,Wu:2016gas}. Taking the assumption $\frac{C_{s\bar{s}}}{C_{ss}}=\frac{C_{c\bar{c}}}{C_{cc}}=\frac{C_{n\bar{n}}}{C_{nn}}$, we obtain $C_{s\bar{s}}=9.4\ \text{MeV}$ and $C_{cc}=3.2\ \text{MeV}$. The effective quark mass gap $\Delta_{sn}=90.6\ \text{MeV}$ has been determined in our previous studies \cite{Cheng:2020nho,Wu:2018xdi}. We continue to use this value in the pentaquark case. The measured mass of the reference $P_\psi^N(4312)$ is 4311.9 MeV \cite{ParticleDataGroup:2024cfk} while the calculated CMI eigenvalue is $(E_{CMI})_{P_\psi^N(4312)}=-70.7\ \text{MeV}$ \cite{Li:2023aui}. Thus, one has $\tilde{m}_{penta}=4382.6$ MeV.

To give upper limits for the pentaquark masses using Eq. \eqref{mass1}, we also need to know the effective quark masses. Their extracted values from conventional hadrons are $m_n=361.8\ \text{MeV}$, $m_s=542.4\ \text{MeV}$, and $m_c=1724.1\ \text{MeV}$ \cite{Li:2023wug}. To give pentaquark masses using Eq. \eqref{mass2} and to calculate decay widths, one employs the following hadron masses \cite{ParticleDataGroup:2024cfk}: $M_{\Lambda_c}=2286.5\ \text{MeV}$, $M_{\Sigma_c}=2453.5\ \text{MeV}$, $M_{\Sigma_c^*}=2518.1\ \text{MeV}$, $M_{\Xi_{cc}}=3621.6\ \text{MeV}$, $M_{\Omega_c}=2695.2\ \text{MeV}$, $M_{\Omega_c^*}=2765.9\ \text{MeV}$, $M_{\Xi_{c}}=2469.1\ \text{MeV}$, $M_{\Xi_c'}=2578.5\ \text{MeV}$, $M_{\Xi_c^*}=2645.6\ \text{MeV}$, $M_D=1867.3\ \text{MeV}$, $M_{D^*}=2008.6\ \text{MeV}$, $M_{\pi}=137.3\ \text{MeV}$, $M_{\rho}=775.3\ \text{MeV}$, $M_{\omega}=782.7\ \text{MeV}$, $M_{D_s}=1968.4\ \text{MeV}$, $M_{D_s^*}=2112.2\ \text{MeV}$, $M_K=495.6\ \text{MeV}$, $M_{K^*}=895.5\ \text{MeV}$, $M_{\phi}=1019.5\ \text{MeV}$, $M_{\eta}=547.9\ \text{MeV}$, and $M_{\eta'}=957.8\ \text{MeV}$. The baryons $\Xi_{cc}^*$, $\Omega_{cc}$, and $\Omega_{cc}^*$ are yet to be discovered. To estimate their masses, we employ the CMI model. The resulting values are $M_{\Xi_{cc}^*}=M_{\Xi_{cc}}+16C_{cn}=3685.6\ \text{MeV}$, $M_{\Omega_{cc}}=M_{\Xi_{cc}}+\frac{32}{3}(C_{cn}-C_{cs})+\Delta_{sn}=3709.0\ \text{MeV}$, and $M_{\Omega_{cc}^*}=M_{\Omega_{cc}}+16C_{cs}=3777.8\ \text{MeV}$. The mass difference between $\Omega_{cc}^*$ and $\Xi_{cc}^*$ is consistent with the heavy diquark-antiquark symmetry \cite{Cheng:2020wxa}.

The rearrangement decay widths depend on the constant ${\cal C}$ which should be different from system to system. Since no doubly-charmed pentaquark state is observed yet, one cannot determine the value of ${\cal C}$ directly. Noticing that $P_\psi^N$ states also contain two charm quark components and three light quark components, the ${\cal C}$ values for $c\bar{c}qqq$ and $ccqq\bar{q}$ systems may be not so different. At present, we estimate the $ccqq\bar{q}$ widths with ${\cal C}$ extracted from $\Gamma_{P_\psi^N(4312)}=9.8$ MeV \cite{ParticleDataGroup:2024cfk}. With the assumption $\Gamma_{total}=\Gamma_{sum}$, what we get is ${\cal C}=4647.9\ \text{MeV}$ \cite{Li:2023aui}.

Utilizing the aforementioned parameters, we can derive the mass spectra and rearrangement decays of $ccqq\bar{q}$ compact pentaquark states. We classify the studied systems into eight cases, $(ccnn\bar{n})^{I_{nn}=1}$, $(ccnn\bar{n})^{I_{nn}=0}$, $(ccnn\bar{s})^{I_{nn}=1}$, $(ccnn\bar{s})^{I_{nn}=0}$, $ccss\bar{n}$, $ccss\bar{s}$, $ccns\bar{n}$, and $ccns\bar{s}$. It should be noted that the $(ccnn\bar{n})^{I_{nn}=1}$ case has degenerate $I=\frac32$ and $\frac12$ states while the $ccns\bar{n}$ case has degenerate $I=1$ and 0 states since the isospin breaking effects are not considered. For decays of degenerate states with different isospins but same spin, the partial widths of the same channel may be different due to their flavor wavefunctions, which can be seen in the following discussions.

\begin{figure}[htbp]\centering
\begin{tabular}{ccc}	\includegraphics[width=0.4\textwidth]{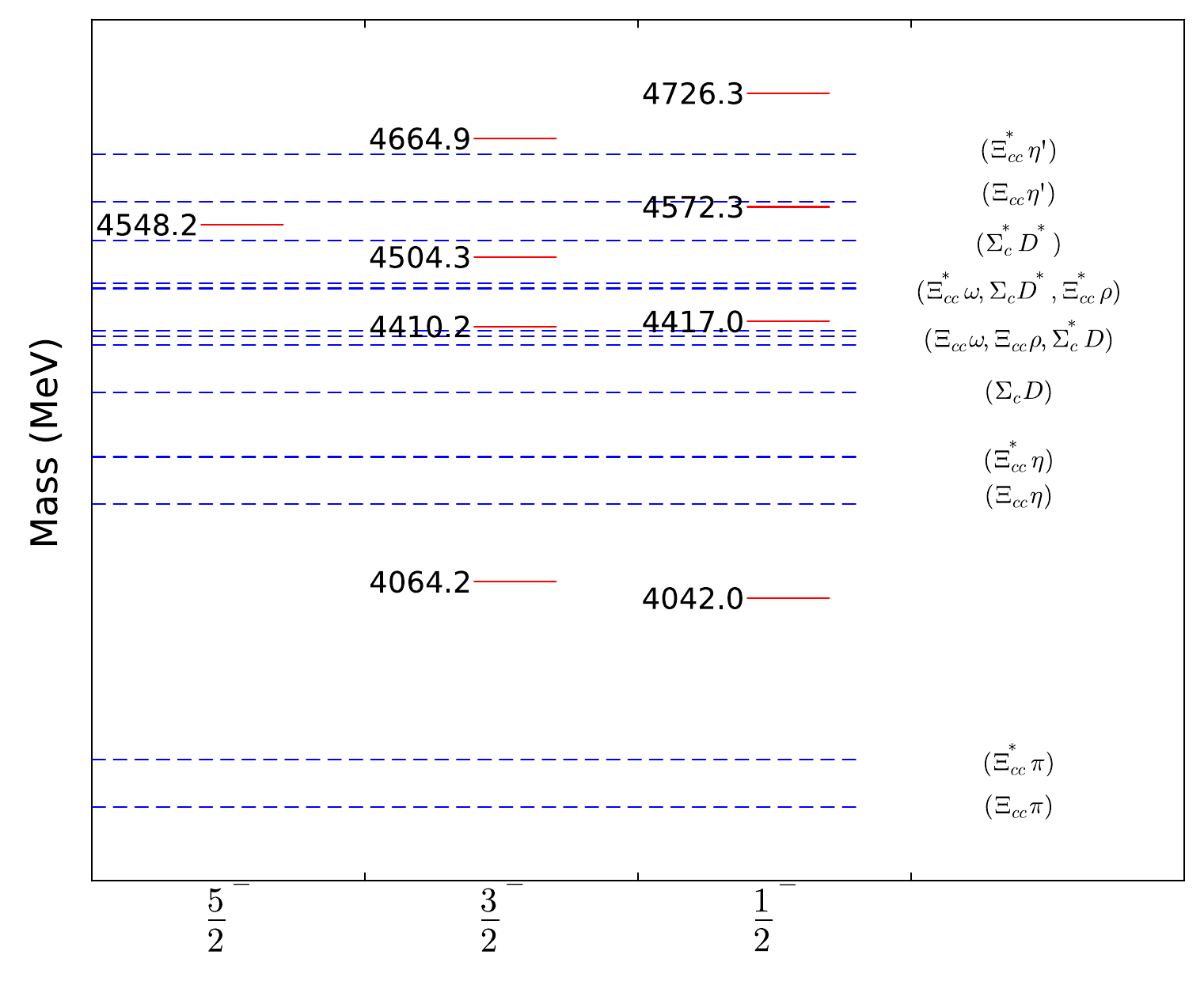}&$\qquad$&\includegraphics[width=0.4\textwidth]{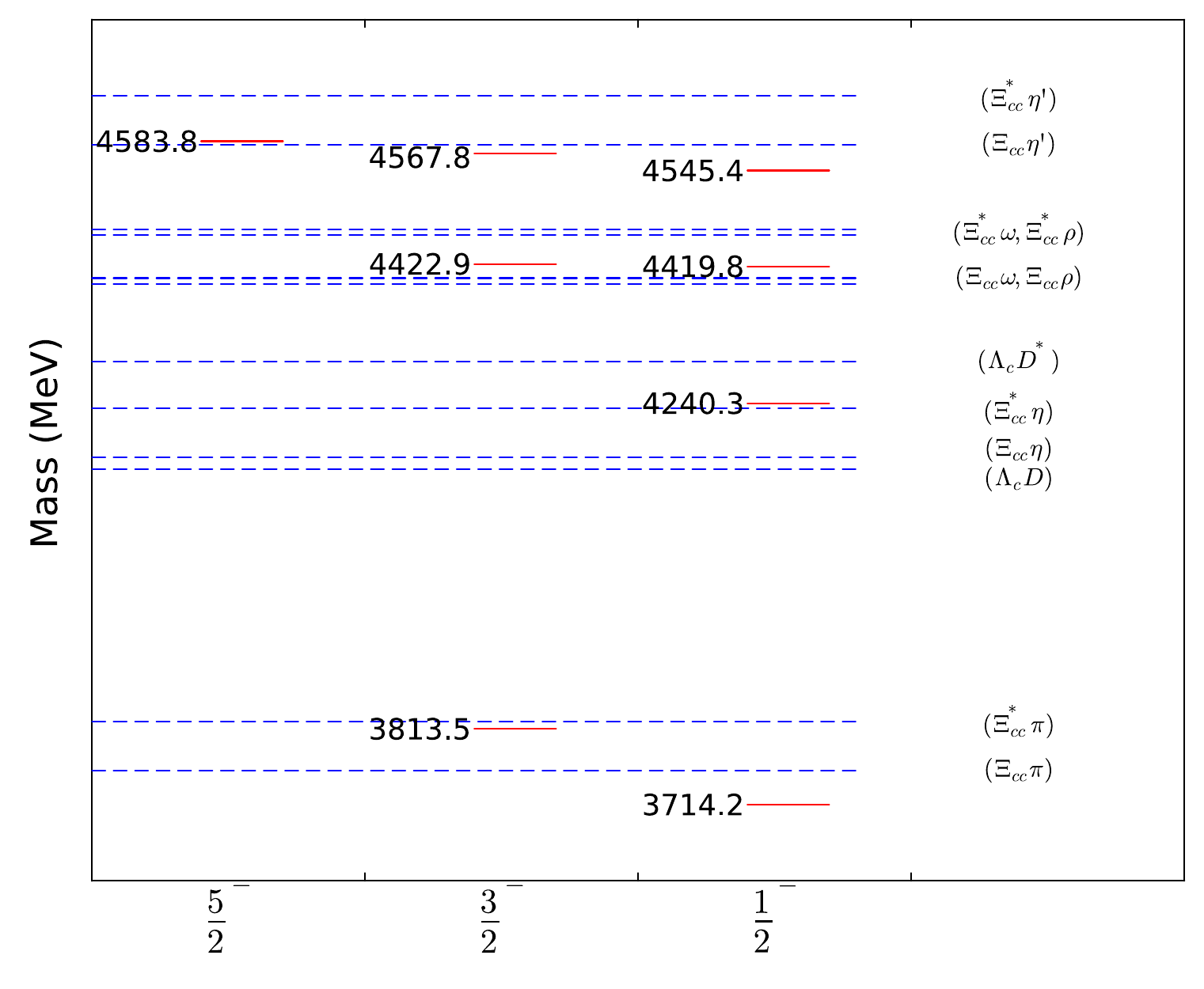}\\
(a) $ccnn\bar{n}$ states with $I_{nn}=1$ and $I=\frac32,\frac12$ & $\qquad$& (b) $ccnn\bar{n}$ states with $I_{nn}=0$ and $I=\frac12$\\
\includegraphics[width=0.4\textwidth]{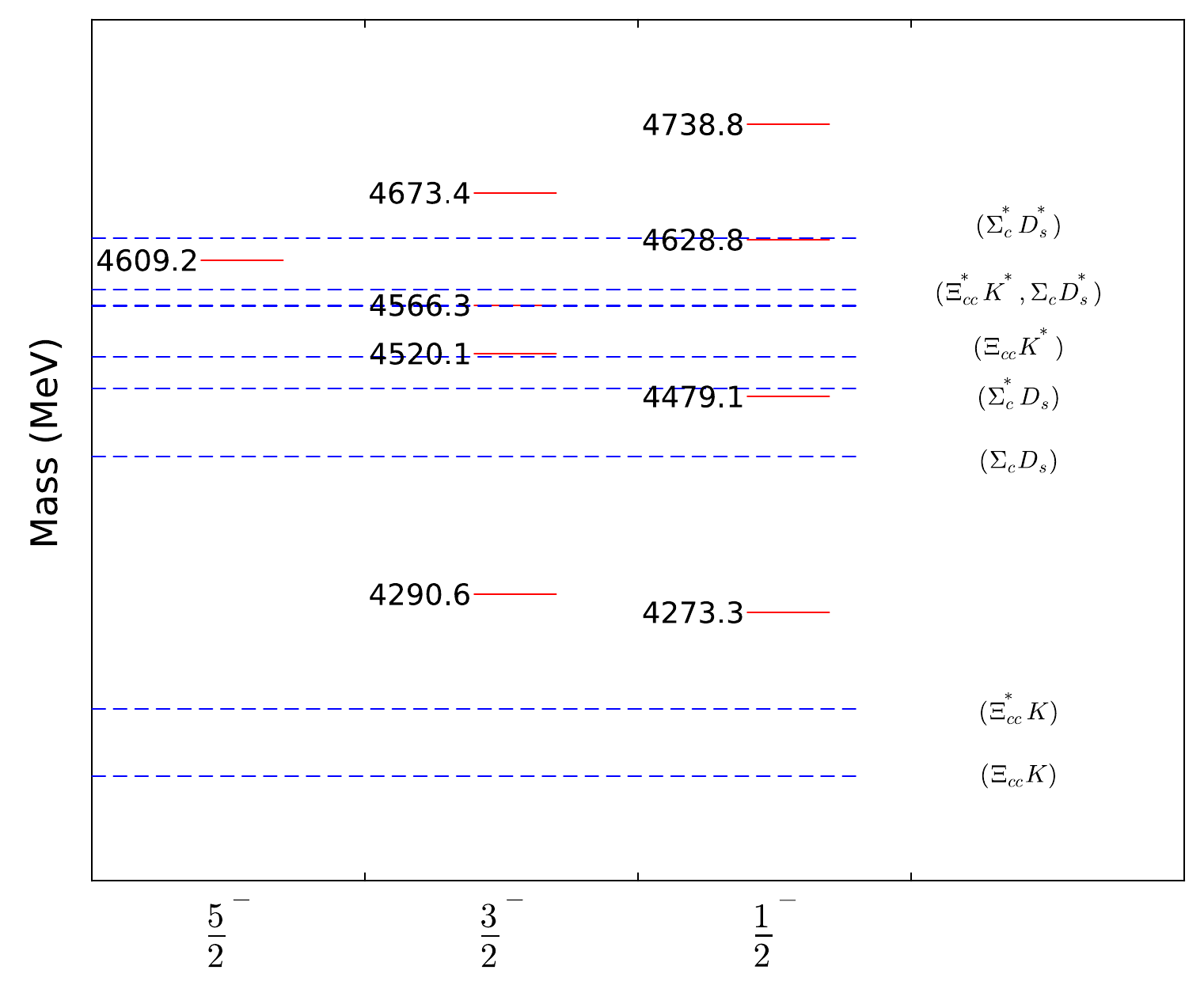}&$\qquad$&\includegraphics[width=0.4\textwidth]{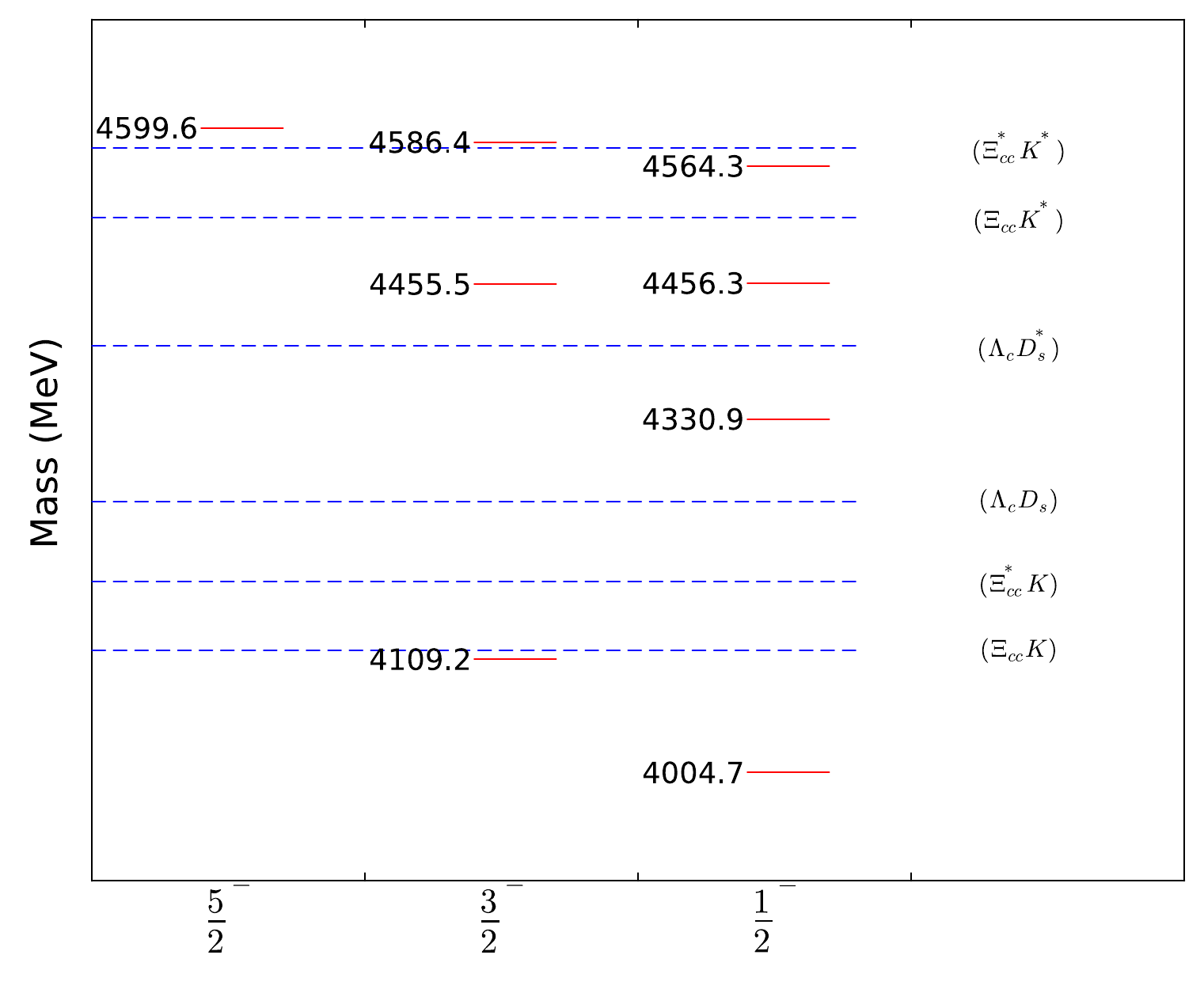}\\
(c) $ccnn\bar{s}$ states with $I=1$ & $\qquad$& (d) $ccnn\bar{s}$ states with $I=0$\\
\includegraphics[width=0.4\textwidth]{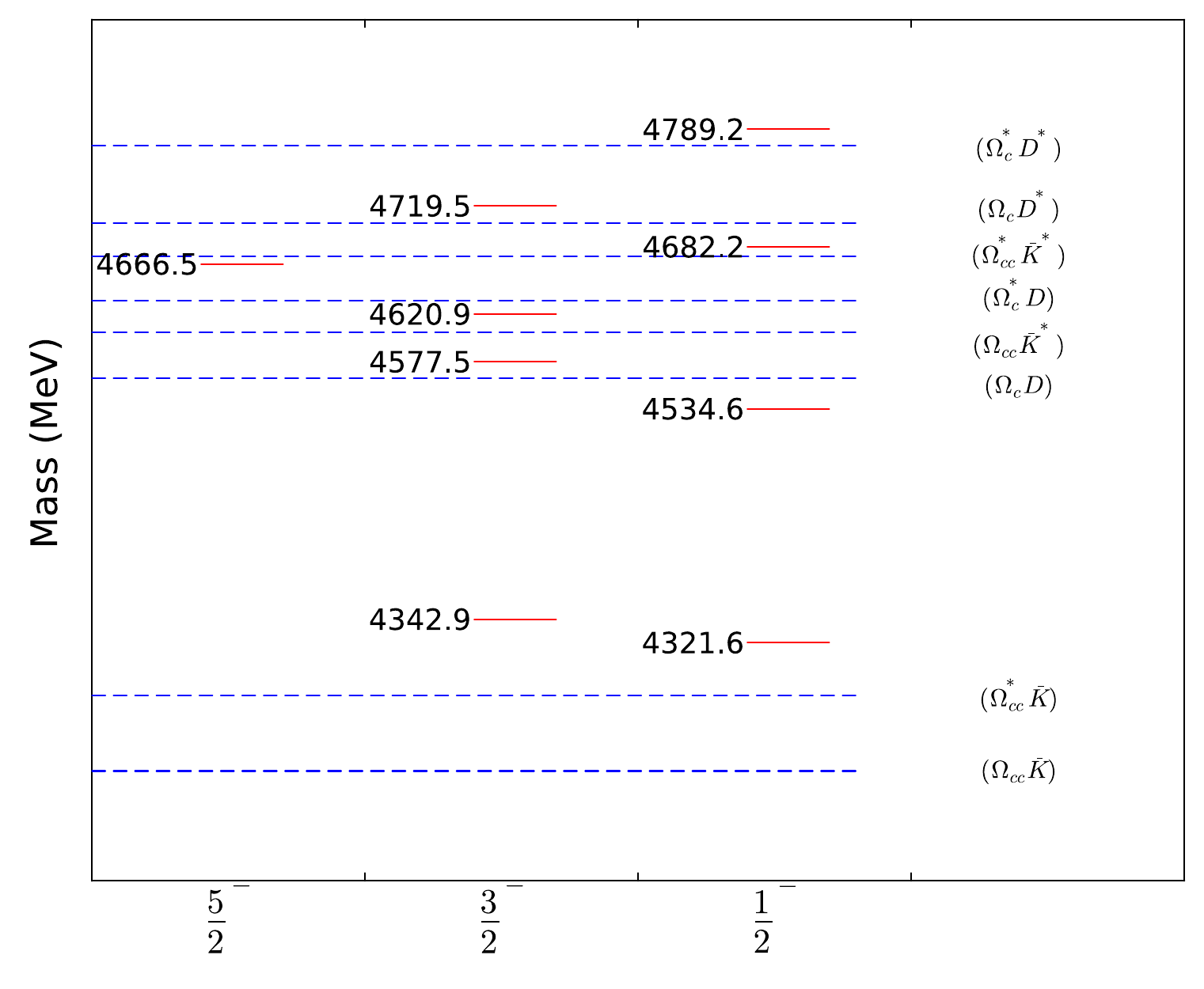}&$\qquad$&\includegraphics[width=0.4\textwidth]{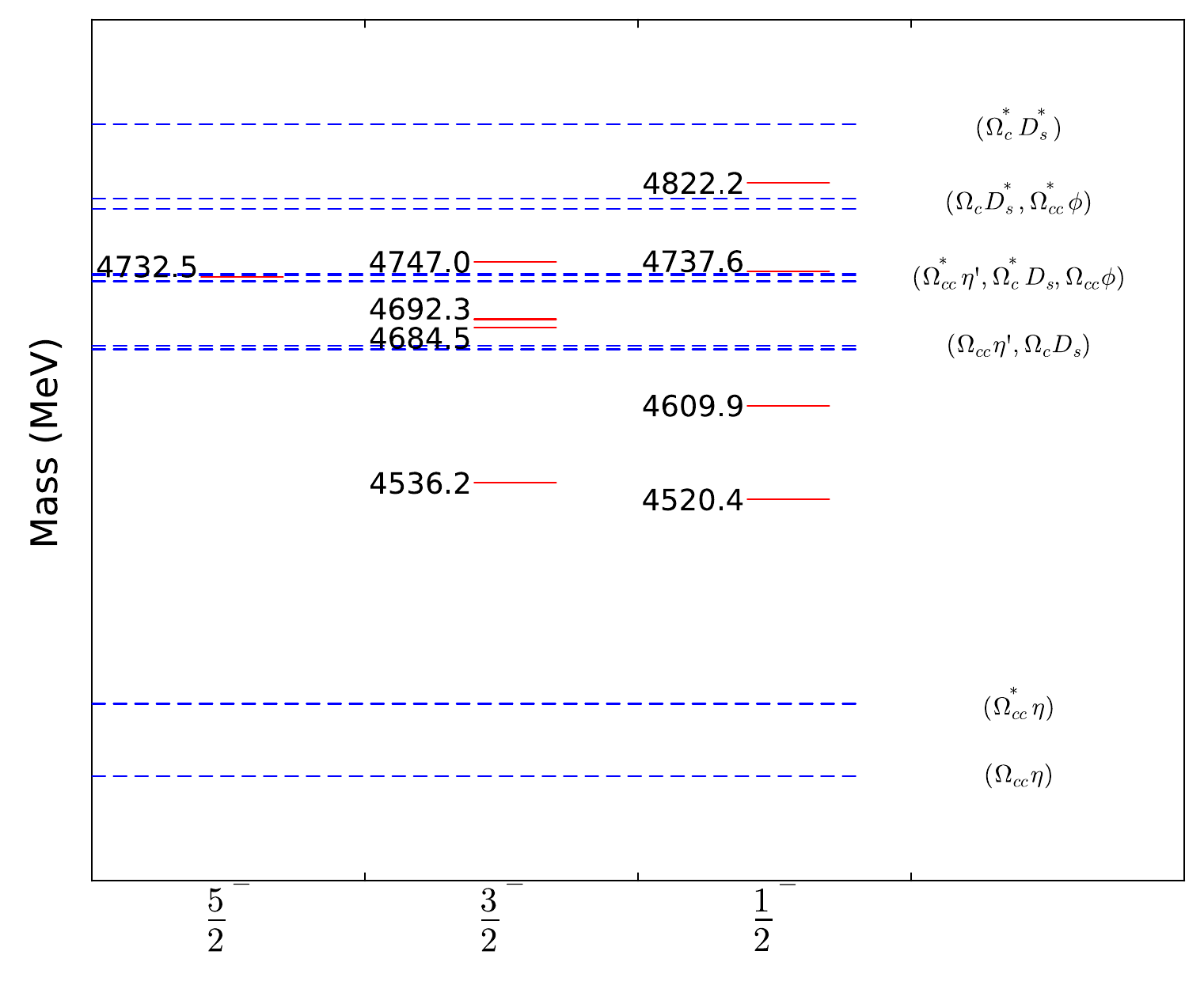}\\
(e) $ccss\bar{n}$ states & $\qquad$& (f) $ccss\bar{s}$ states\\
\includegraphics[width=0.4\textwidth]{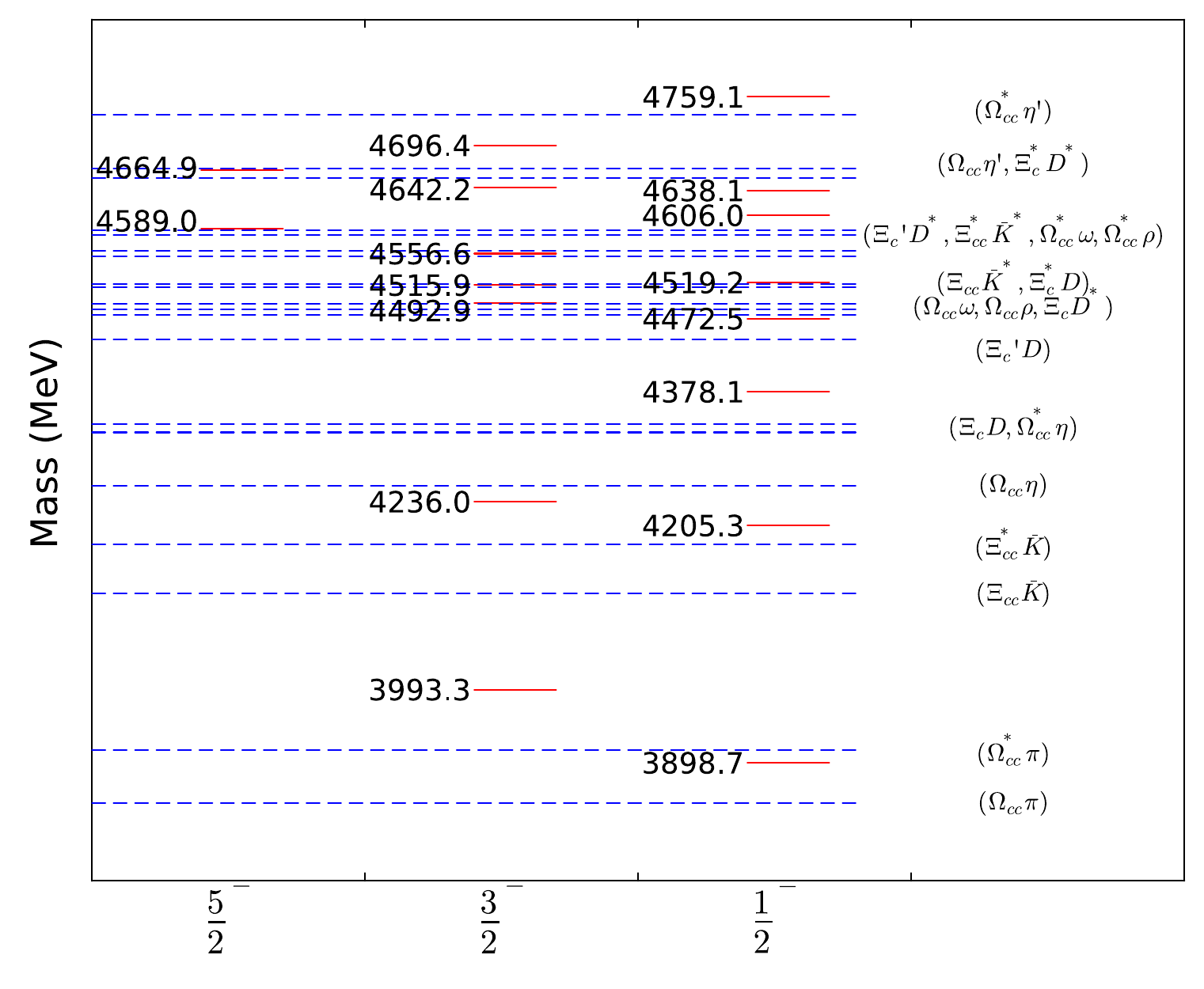}&$\qquad$&\includegraphics[width=0.4\textwidth]{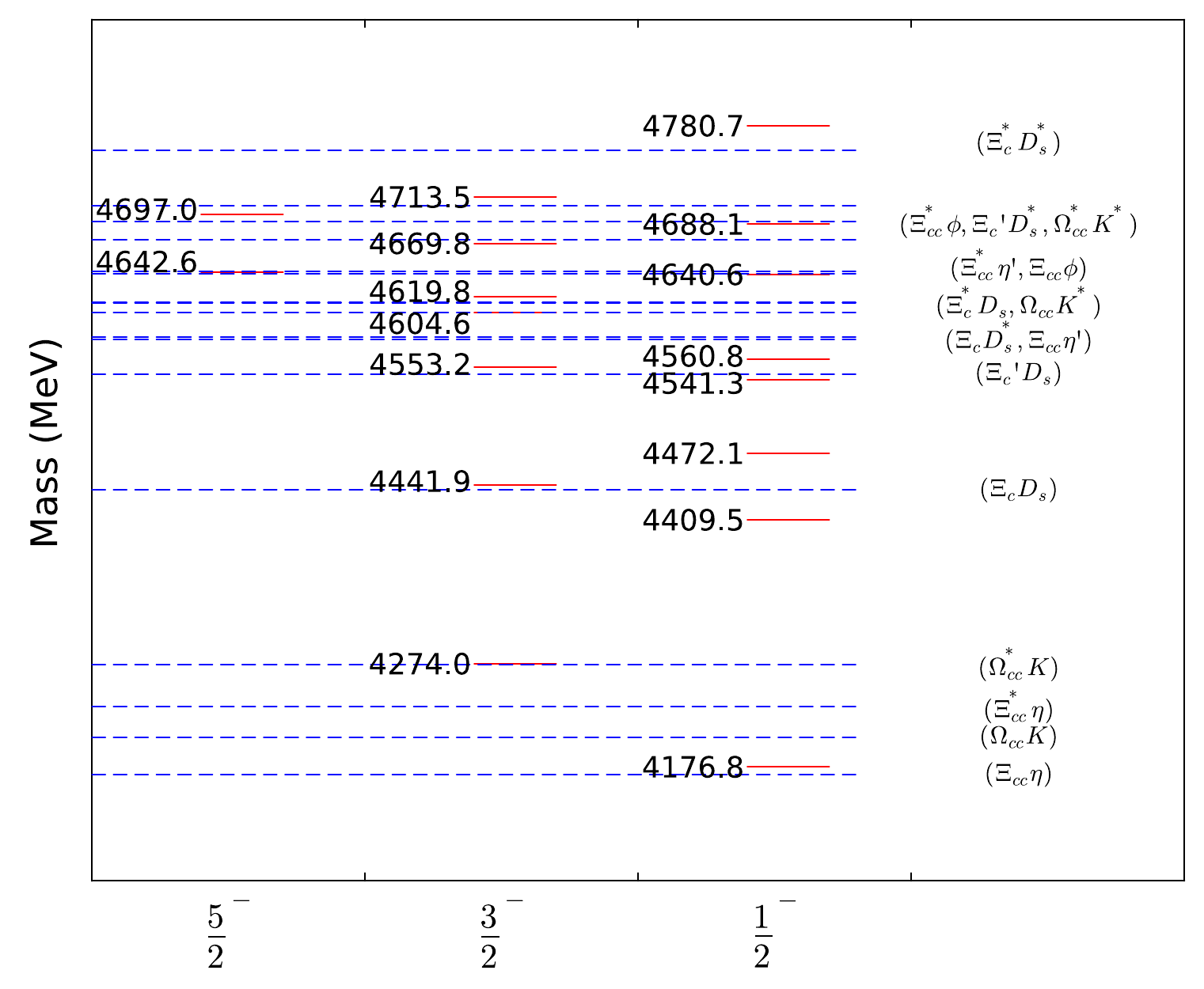}\\
(g) $ccns\bar{n}$ states with $I=1,0$ & $\qquad$& (h) $ccns\bar{s}$ states
\end{tabular}
\caption{Relative positions for the doubly-charmed pentaquark states and corresponding baryon-meson thresholds.}\label{figures}
\end{figure}
	
\subsection{Spectra and decay widths of $ccnn\bar{n}$ states}

\begin{table}[htbp]\caption{Numerical results for the $ccnn\bar{n}$ states. The CMI matrix $\langle H_{CMI}\rangle$, its eigenvalues, and pentaquark masses using Eq. \eqref{mass3detail} are listed in the second, third, and fifth columns, respectively, in units of MeV. The estimated masses using Eq. \eqref{mass2} with two different baryon-meson thresholds are given in the sixth and seven columns and those using Eq. \eqref{mass1} are shown in the last column. The corresponding base vectors of $\langle H_{CMI}\rangle$ can be found in table \ref{basis}.}\label{massofccnnn}
   \resizebox{\linewidth}{!}{
\begin{tabular}{cccccccc}\hline\hline
\multicolumn{8}{c}{$ccnn\bar{n}$ states with $I_{nn}=1$ and $I=\frac32,\frac12$}\\\hline
$J^P$&$\langle H_{CMI}\rangle$&Eigenvalue&Eigenvector&Mass&$\Sigma_cD$&$\Xi_{cc}\pi$&Upper limits\\\hline
$\frac52^-$&$\left(\begin{array}{c}165.6\end{array}\right)$&$\left(\begin{array}{c}165.6\end{array}\right)$&$\left[\begin{array}{c}\{1.00\}\end{array}\right]$&$\left(\begin{array}{c}4548.2\end{array}\right)$&$\left(\begin{array}{c}4585.6\end{array}\right)$&$\left(\begin{array}{c}4437.0\end{array}\right)$&$\left(\begin{array}{c}4699.2\end{array}\right)$\\$\frac32^-$&$\left(\begin{array}{cccc}-77.7&69.5&267.4&59.0\\69.5&95.6&-87.6&-5.6\\267.4&-87.6&73.3&-16.0\\59.0&-5.6&-16.0&22.0\end{array}\right)$&$\left(\begin{array}{c}282.3\\121.7\\27.6\\-318.4\end{array}\right)$&$\left[\begin{array}{cccc}\{-0.57,0.16,-0.80,-0.08\}\\\{-0.31,-0.94,0.04,-0.14\}\\\{0.03,-0.16,-0.16,0.97\}\\\{0.76,-0.25,-0.58,-0.16\}\end{array}\right]$&$\left(\begin{array}{c}4664.9\\4504.3\\4410.2\\4064.2\end{array}\right)$&$\left(\begin{array}{c}4702.3\\4541.7\\4447.6\\4101.6\end{array}\right)$&$\left(\begin{array}{c}4553.7\\4393.1\\4299.0\\3953.1\end{array}\right)$&$\left(\begin{array}{c}4815.9\\4655.3\\4561.2\\4215.2\end{array}\right)$\\$\frac12^-$&$\left(\begin{array}{cccc}36.3&87.9&-169.1&-37.3\\87.9&-50.4&271.2&-84.8\\-169.1&271.2&99.7&-16.0\\-37.3&-84.8&-16.0&141.6\end{array}\right)$&$\left(\begin{array}{c}343.7\\189.7\\34.4\\-340.6\end{array}\right)$&$\left[\begin{array}{cccc}\{-0.25,0.53,0.78,-0.24\}\\\{-0.56,-0.23,0.22,0.77\}\\\{-0.67,-0.43,-0.10,-0.59\}\\\{0.42,-0.69,0.58,-0.07\}\end{array}\right]$&$\left(\begin{array}{c}4726.3\\4572.3\\4417.0\\4042.0\end{array}\right)$&$\left(\begin{array}{c}4763.7\\4609.7\\4454.4\\4079.4\end{array}\right)$&$\left(\begin{array}{c}4615.1\\4461.1\\4305.8\\3930.9\end{array}\right)$&$\left(\begin{array}{c}4877.3\\4723.3\\4568.0\\4193.0\end{array}\right)$\\\hline
\multicolumn{8}{c}{$ccnn\bar{n}$ states with $I_{nn}=0$ and $I=\frac12$}\\\hline
$J^P$&$\langle H_{CMI}\rangle$&Eigenvalue&Eigenvector&Mass&$\Sigma_cD$&$\Xi_{cc}\pi$&Upper limits\\\hline
$\frac52^-$&$\left(\begin{array}{c}201.2\end{array}\right)$&$\left(\begin{array}{c}201.2\end{array}\right)$&$\left[\begin{array}{c}\{1.00\}\end{array}\right]$&$\left(\begin{array}{c}4583.8\end{array}\right)$&$\left(\begin{array}{c}4621.2\end{array}\right)$&$\left(\begin{array}{c}4472.6\end{array}\right)$&$\left(\begin{array}{c}4734.8\end{array}\right)$\\$\frac32^-$&$\left(\begin{array}{ccc}-275.1&232.7&267.4\\232.7&52.6&-87.6\\267.4&-87.6&-121.1\end{array}\right)$&$\left(\begin{array}{c}185.2\\40.3\\-569.1\end{array}\right)$&$\left[\begin{array}{ccc}\{0.55,0.80,0.25\}\\\{0.33,-0.48,0.81\}\\\{0.77,-0.36,-0.53\}\end{array}\right]$&$\left(\begin{array}{c}4567.8\\4422.9\\3813.5\end{array}\right)$&$\left(\begin{array}{c}4605.2\\4460.3\\3850.9\end{array}\right)$&$\left(\begin{array}{c}4456.7\\4311.7\\3702.3\end{array}\right)$&$\left(\begin{array}{c}4718.8\\4573.9\\3964.5\end{array}\right)$\\$\frac12^-$&$\left(\begin{array}{cccc}-69.3&27.7&294.3&-169.1\\27.7&-134.4&0.0&64.7\\294.3&0.0&-233.2&271.2\\-169.1&64.7&271.2&-173.9\end{array}\right)$&$\left(\begin{array}{c}162.8\\37.2\\-142.3\\-668.4\end{array}\right)$&$\left[\begin{array}{cccc}\{-0.71,-0.11,-0.67,-0.20\}\\\{-0.50,0.22,0.26,0.80\}\\\{-0.08,-0.96,0.20,0.15\}\\\{0.49,-0.09,-0.67,0.55\}\end{array}\right]$&$\left(\begin{array}{c}4545.4\\4419.8\\4240.3\\3714.2\end{array}\right)$&$\left(\begin{array}{c}4582.8\\4457.2\\4277.7\\3751.6\end{array}\right)$&$\left(\begin{array}{c}4434.3\\4308.6\\4129.1\\3603.0\end{array}\right)$&$\left(\begin{array}{c}4696.4\\4570.8\\4391.3\\3865.2\end{array}\right)$\\\hline\hline\end{tabular}}\end{table}

\begin{table}[htbp]\caption{Rearrangement decays for the $ccnn\bar{n}$ states. The numbers in the parentheses are $(100|\mathcal{M}|^2/{\cal C}^2,\ \Gamma_i)$. The pentaquark masses, their partial widths ($\Gamma_i$'s) and total width ($\Gamma_{sum}$) are given in units of MeV. The symbol `-' means that the corresponding channel is kinematically forbidden.}\label{decayofccnnn}
\resizebox{\linewidth}{!}{
\begin{tabular}{ccccccccccccccc}\hline\hline
$J^P$&Mass&\multicolumn{12}{c}{Channels}&$\Gamma_{sum}$\\\hline\multicolumn{15}{c}{$ccnn\bar{n}$ states with $I_{nn}=1$ and $I=\frac32$}\\\hline
&&$\Sigma_c^*D^*$&$\Xi_{cc}^*\rho$&&&&&&&&&&&\\
$\frac52^-$&$\left[\begin{array}{c}4548.2\end{array}\right]$&$\left[\begin{array}{c}(33.3,6.0)\end{array}\right]$&$\left[\begin{array}{c}(33.3,9.4)\end{array}\right]$&&&&&&&&&&&$\left[\begin{array}{c}15.4\end{array}\right]$\\
&&$\Sigma_c^*D^*$&$\Sigma_cD^*$&$\Sigma_c^*D$&$\Xi_{cc}^*\rho$&$\Xi_{cc}^*\pi$&$\Xi_{cc}\rho$&&&&&&&\\
$\frac32^-$&$\left[\begin{array}{c}4664.9\\4504.3\\4410.2\\4064.2\end{array}\right]$&$\left[\begin{array}{c}(8.5,3.8)\\(0.4,-)\\(32.1,-)\\(6.2,-)\end{array}\right]$&$\left[\begin{array}{c}(0.1,0.0)\\(25.8,6.6)\\(14.7,-)\\(3.8,-)\end{array}\right]$&$\left[\begin{array}{c}(3.8,2.4)\\(7.7,3.4)\\(17.2,3.6)\\(13.0,-)\end{array}\right]$&$\left[\begin{array}{c}(33.7,14.2)\\(11.1,2.2)\\(1.3,-)\\(1.1,-)\end{array}\right]$&$\left[\begin{array}{c}(0.1,0.0)\\(0.1,0.0)\\(1.3,0.8)\\(40.2,14.0)\end{array}\right]$&$\left[\begin{array}{c}(20.5,10.0)\\(21.6,7.0)\\(0.0,0.0)\\(2.4,-)\end{array}\right]$&&&&&&&$\left[\begin{array}{c}30.4\\19.2\\4.4\\14.0\end{array}\right]$\\&&$\Sigma_c^*D^*$&$\Sigma_cD^*$&$\Sigma_cD$&$\Xi_{cc}^*\rho$&$\Xi_{cc}\pi$&$\Xi_{cc}\rho$&&&&&&&\\
$\frac12^-$&$\left[\begin{array}{c}4726.3\\4572.3\\4417.0\\4042.0\end{array}\right]$&$\left[\begin{array}{c}(11.3,5.8)\\(34.6,9.2)\\(6.3,-)\\(3.4,-)\end{array}\right]$&$\left[\begin{array}{c}(3.7,2.2)\\(16.4,6.8)\\(0.9,-)\\(15.1,-)\end{array}\right]$&$\left[\begin{array}{c}(0.2,0.2)\\(0.3,0.2)\\(37.4,15.0)\\(3.7,-)\end{array}\right]$&$\left[\begin{array}{c}(48.1,22.8)\\(0.5,0.2)\\(4.5,-)\\(2.5,-)\end{array}\right]$&$\left[\begin{array}{c}(0.1,0.0)\\(0.0,0.0)\\(0.4,0.2)\\(41.2,16.4)\end{array}\right]$&$\left[\begin{array}{c}(3.3,1.8)\\(14.8,6.0)\\(17.2,2.4)\\(0.9,-)\end{array}\right]$&&&&&&&$\left[\begin{array}{c}32.8\\22.4\\17.6\\16.4\end{array}\right]$\\\hline
\multicolumn{15}{c}{$ccnn\bar{n}$ states with $I_{nn}=1$ and $I=\frac12$}\\\hline
&&$\Sigma_c^*D^*$&$\Xi_{cc}^*\rho$&$\Xi_{cc}^*\omega$&&&&&&&&&&\\
$\frac52^-$&$\left[\begin{array}{c}4548.2\end{array}\right]$&$\left[\begin{array}{c}(33.3,6.0)\end{array}\right]$&$\left[\begin{array}{c}(8.3,2.4)\end{array}\right]$&$\left[\begin{array}{c}(25.0,6.8)\end{array}\right]$&&&&&&&&&&$\left[\begin{array}{c}15.1\end{array}\right]$\\
&&$\Sigma_c^*D^*$&$\Sigma_cD^*$&$\Sigma_c^*D$&$\Xi_{cc}^*\rho$&$\Xi_{cc}^*\pi$&$\Xi_{cc}\rho$&$\Xi_{cc}^*\omega$&$\Xi_{cc}\omega$&$\Xi_{cc}^*\eta$&$\Xi_{cc}^*\eta'$&&&\\
$\frac32^-$&$\left[\begin{array}{c}4664.9\\4504.3\\4410.2\\4064.2\end{array}\right]$&$\left[\begin{array}{c}(8.5,3.8)\\(0.4,-)\\(32.1,-)\\(6.2,-)\end{array}\right]$&$\left[\begin{array}{c}(0.1,0.0)\\(25.8,6.6)\\(14.7,-)\\(3.8,-)\end{array}\right]$&$\left[\begin{array}{c}(3.8,2.4)\\(7.7,3.4)\\(17.2,3.6)\\(13.0,-)\end{array}\right]$&$\left[\begin{array}{c}(8.4,3.6)\\(2.8,0.6)\\(0.3,-)\\(0.3,-)\end{array}\right]$&$\left[\begin{array}{c}(0.0,0.0)\\(0.0,0.0)\\(0.3,0.2)\\(10.1,3.5)\end{array}\right]$&$\left[\begin{array}{c}(5.1,2.5)\\(5.4,1.8)\\(0.0,0.0)\\(0.6,-)\end{array}\right]$&$\left[\begin{array}{c}(6.4,2.7)\\(0.3,0.0)\\(24.1,-)\\(4.7,-)\end{array}\right]$&$\left[\begin{array}{c}(15.4,7.4)\\(16.2,5.0)\\(0.0,0.0)\\(1.8,-)\end{array}\right]$&$\left[\begin{array}{c}(0.0,0.0)\\(0.1,0.0)\\(0.5,0.2)\\(15.9,-)\end{array}\right]$&$\left[\begin{array}{c}(0.0,0.0)\\(0.1,-)\\(0.5,-)\\(14.3,-)\end{array}\right]$&&&$\left[\begin{array}{c}22.3\\17.3\\4.0\\3.5\end{array}\right]$\\
&&$\Sigma_c^*D^*$&$\Sigma_cD^*$&$\Sigma_cD$&$\Xi_{cc}^*\rho$&$\Xi_{cc}\pi$&$\Xi_{cc}\rho$&$\Xi_{cc}^*\omega$&$\Xi_{cc}\omega$&$\Xi_{cc}\eta$&$\Xi_{cc}\eta'$&&&\\
$\frac12^-$&$\left[\begin{array}{c}4726.3\\4572.3\\4417.0\\4042.0\end{array}\right]$&$\left[\begin{array}{c}(11.3,5.8)\\(34.6,9.2)\\(6.3,-)\\(3.4,-)\end{array}\right]$&$\left[\begin{array}{c}(3.7,2.2)\\(16.4,6.8)\\(0.9,-)\\(15.1,-)\end{array}\right]$&$\left[\begin{array}{c}(0.2,0.2)\\(0.3,0.2)\\(37.4,15.0)\\(3.7,-)\end{array}\right]$&$\left[\begin{array}{c}(12.0,5.7)\\(0.1,0.0)\\(1.1,-)\\(0.6,-)\end{array}\right]$&$\left[\begin{array}{c}(0.0,0.0)\\(0.0,0.0)\\(0.1,0.1)\\(10.3,4.1)\end{array}\right]$&$\left[\begin{array}{c}(0.8,0.5)\\(3.7,1.5)\\(4.3,0.6)\\(0.2,-)\end{array}\right]$&$\left[\begin{array}{c}(8.5,3.9)\\(26.0,8.0)\\(4.7,-)\\(2.6,-)\end{array}\right]$&$\left[\begin{array}{c}(2.5,1.4)\\(11.1,4.4)\\(12.9,1.5)\\(0.7,-)\end{array}\right]$&$\left[\begin{array}{c}(0.0,0.0)\\(0.0,0.0)\\(0.2,0.0)\\(16.3,-)\end{array}\right]$&$\left[\begin{array}{c}(0.0,0.0)\\(0.0,-)\\(0.2,-)\\(14.6,-)\end{array}\right]$&&&$\left[\begin{array}{c}19.6\\30.1\\17.2\\4.1\end{array}\right]$\\\hline
\multicolumn{15}{c}{$ccnn\bar{n}$ states with $I_{nn}=0$ and $I=\frac12$}\\\hline
&&&$\Xi_{cc}^*\rho$&$\Xi_{cc}^*\omega$&&&&&&&&&&\\
$\frac52^-$&$\left[\begin{array}{c}4583.8\end{array}\right]$&&$\left[\begin{array}{c}(50.0,16.7)\end{array}\right]$&$\left[\begin{array}{c}(16.7,5.4)\end{array}\right]$&&&&&&&&&&$\left[\begin{array}{c}22.1\end{array}\right]$\\
&&&$\Lambda_cD^*$&&$\Xi_{cc}^*\rho$&$\Xi_{cc}^*\pi$&$\Xi_{cc}\rho$
&$\Xi_{cc}^*\omega$&$\Xi_{cc}\omega$&$\Xi_{cc}^*\eta$&$\Xi_{cc}^*\eta'$&&&\\
$\frac32^-$&$\left[\begin{array}{c}4567.8\\4422.9\\3813.5\end{array}\right]$&&$\left[\begin{array}{c}(2.1,1.4)\\(21.9,10.2)\\(9.3,-)\end{array}\right]$&&$\left[\begin{array}{c}(33.2,10.4)\\(6.2,-)\\(0.2,-)\end{array}\right]$&$\left[\begin{array}{c}(0.2,0.2)\\(1.1,0.6)\\(42.5,-)\end{array}\right]$&$\left[\begin{array}{c}(15.0,6.0)\\(26.3,4.2)\\(0.3,-)\end{array}\right]$
&$\left[\begin{array}{c}(0.0,0.0)\\(0.0,-)\\(0.0,-)\end{array}\right]$
&$\left[\begin{array}{c}(5.0,2.0)\\(8.8,1.2)\\(0.1,-)\end{array}\right]$&$\left[\begin{array}{c}(0.0,0.0)\\(0.2,0.1)\\(7.5,-)\end{array}\right]$&$\left[\begin{array}{c}(0.0,-)\\(0.2,-)\\(6.7,-)\end{array}\right]$&&&$\left[\begin{array}{c}19.9\\16.3\\-\end{array}\right]$\\
&&&$\Lambda_cD^*$&$\Lambda_cD$&$\Xi_{cc}^*\rho$&$\Xi_{cc}\pi$&$\Xi_{cc}\rho$&$\Xi_{cc}^*\omega$&$\Xi_{cc}\omega$&$\Xi_{cc}\eta$&$\Xi_{cc}\eta'$&&&\\
$\frac12^-$&$\left[\begin{array}{c}4545.4\\4419.8\\4240.3\\3714.2\end{array}\right]$&&$\left[\begin{array}{c}(1.8,1.2)\\(14.9,6.8)\\(40.7,-)\\(0.9,-)\end{array}\right]$&$\left[\begin{array}{c}(0.3,0.2)\\(9.6,6.4)\\(22.1,9.0)\\(9.7,-)\end{array}\right]$&$\left[\begin{array}{c}(12.5,3.5)\\(18.5,-)\\(2.0,-)\\(0.2,-)\end{array}\right]$&$\left[\begin{array}{c}(0.1,0.0)\\(1.1,0.8)\\(0.8,0.5)\\(41.8,-)\end{array}\right]$&$\left[\begin{array}{c}(35.9,13.4)\\(12.0,1.8)\\(0.0,-)\\(0.1,-)\end{array}\right]$
&$\left[\begin{array}{c}(0.0,0.0)\\(0.0,-)\\(0.0,-)\\(0.0,-)\end{array}\right]$&$\left[\begin{array}{c}(12.0,4.4)\\(4.0,0.5)\\(0.0,-)\\(0.0,-)\end{array}\right]$&$\left[\begin{array}{c}(0.1,0.0)\\(0.6,0.3)\\(0.5,0.2)\\(22.0,-)\end{array}\right]$&$\left[\begin{array}{c}(0.1,-)\\(0.5,-)\\(0.4,-)\\(19.7,-)\end{array}\right]$&&&$\left[\begin{array}{c}22.6\\16.6\\9.6\\-\end{array}\right]$\\\hline\hline
\end{tabular}}
\end{table}
	
The results for the $ccnn\bar{n}$ states are detailed in tables \ref{massofccnnn} and \ref{decayofccnnn}. We present in table \ref{massofccnnn} the CMI matrices, their corresponding eigenvalues and eigenvectors, predicted masses using Eq. \eqref{mass3detail}, estimated values utilizing reference systems $\Sigma_cD$ and $\Xi_{cc}\pi$, and theoretical upper limits for pentaquark masses using Eq. \eqref{mass1}. Table \ref{decayofccnnn} is for the decay widths. From table \ref{massofccnnn}, the predicted results are about 40 MeV smaller than the estimated masses using the $\Sigma_cD$ threshold, and about 150 MeV smaller than the upper limits. This means that the estimation method with baryon-meson thresholds cannot be naively treated as lower limits for pentaquark masses. The situation is different from the tetraquark case. The $ccnn\bar{n}$ masses obtained with the $\Xi_{cc}\pi$ threshold may still serve as theoretical lower limits, while those with the $\Sigma_cD$ threshold can also be treated as reasonable pentaquark masses. If future experiments may find some near-threshold exotica whose masses are around those given here, it probably indicates that the difference between compact pentaquark structure and meson-baryon molecule structure is not so large for them. We hope future studies may check this conjecture. In the diagrams (a) and (b) of Fig. \ref{figures}, we visually depict the relative positions for the obtained $ccnn\bar{n}$ pentaquark states. 

From the isospin $SU(2)$ symmetry, there are two doublets ($I=1/2$) and one quartet ($I=3/2$). The iso-doublet with $I_{nn}=1$ is degenerate with the iso-quartet [Fig. \ref{figures}(a)]. The lowest state belongs to the $I_{nn}=0$ doublet [Fig. \ref{figures}(b)] and has the quantum numbers $J^P=\frac12^-$. Its mass around 3714 MeV is about 45 MeV below the $\Xi_{cc}\pi$ threshold in our model and it should be stable. If the mass is underestimated, the state might be a narrow near-threshold pentaquark. The second lowest state whose spin is $\frac32$ also belongs to the $I_{nn}=0$ doublet. Although its decay into $\Xi_{cc}^*\pi$ is kinematically forbidden with the obtained mass, it can decay into $\Xi_{cc}\pi$ through $D$-wave interactions. It is also possible that this state is above but not far from the $\Xi_{cc}^*\pi$ threshold. In both possibilities, it should be a narrow near-threshold state. The lowest $J=\frac32$ ($J=\frac12$) pentaquark in the $I_{nn}=1$ iso-doublet is about 250 (330) MeV heavier than the corresponding state in the $I_{nn}=0$ iso-doublet and is not stable. For the iso-doublet pentaquarks, the possibility that they mix with the excited $\Xi_{cc}$ baryons is not excluded. The mixing may lead to physical states different from theoretical expectations, which can be understood from the study of $X(3872)$. For example, if the theoretical mass of a $\frac32^-$ excited $\Xi_{cc}$ is around 3989 MeV \cite{Perez-Rubio:2015zqb} and its mixing with the obtained pentaquark around 4064 or 3814 MeV cannot be ignored, the observation of a $\frac32^-$ $\Xi_{cc}$ below 3989 MeV is possible. The predicted $I=\frac32$ pentaquarks do not mix with the conventional $\Xi_{cc}$ states and are all unstable.

Any allowed two-body channels may be employed to search for the unstable pentaquark states. We move on to their decay properties and first consider the $I_{nn}=1$ ($I=\frac32$ or $I=\frac12$) case. The obtained pentaquarks range from 4042.0 to 4726.3 MeV, with the highest and lowest states  possessing $J^P=\frac12^-$. For two degenerate $I=\frac32$ and $I=\frac12$ pentaquarks, the latter state has more decay modes than the former one, but the obtained width usually smaller. One may understand this feature from the fact that the theoretical total width is determined by the coupling strength between initial and final states, number of decay channels, and phase space together.

There is only one $I(J^P)=\frac32(\frac52^-)$ pentaquark located around 4548 MeV, featuring two open decay channels $\Sigma_c^*D^*$ and $\Xi_{cc}^*\rho$. Its $I=\frac12$ partner has an additional decay channel $\Xi_{cc}^*\omega$. Although the coupling of the $I=\frac12$ state with $\Xi_{cc}^*\rho$ is significantly weaker than that of the $I=\frac32$ state and the partial width smaller, the partial width of the $\Xi_{cc}^*\omega$ channel for the $I=\frac12$ state is considerable. As a result, the total width of the $I=\frac12$ state is nearly equal to that of the $I=\frac32$ state. In reality, the decays also involve more channels ($D$-wave, three-body, etc) and the situation would be more complicated. There are four $I(J^P)=\frac32(\frac32^-)$ pentaquarks ranging from 4064.2 to 4664.9 MeV. The highest state has main decay channels $\Sigma_c^*D^*$, $\Sigma_c^*D$, $\Xi_{cc}^*\rho$, and $\Xi_{cc}\rho$. The second highest has dominant channels $\Sigma_c^*D$, $\Sigma_cD^*$, $\Xi_{cc}^*\rho$, and $\Xi_{cc}\rho$. The second lowest state primarily decays into $\Sigma_c^*D$ and $\Xi_{cc}^*\pi$, and the lowest state decays mainly into $\Xi_{cc}^*\pi$. There are more $S$-wave channels for the $I=\frac12$ partner pentaquarks. The highest state has six main channels: $\Sigma_c^*D^*$, $\Sigma_c^*D$, $\Xi_{cc}^*\rho$, $\Xi_{cc}\rho$, $\Xi_{cc}^*\omega$, and $\Xi_{cc}\omega$. The second highest state features five dominant channels: $\Sigma_c^*D$, $\Sigma_cD^*$, $\Xi_{cc}^*\rho$, $\Xi_{cc}\rho$, and $\Xi_{cc}\omega$. Of the three dominant channels for the second lowest state, the partial width in the $\Sigma_c^*D$ channel is larger than those in the $\Xi_{cc}^*\pi$ and $\Xi_{cc}^*\eta$ channels. The $\Xi_{cc}^*\pi$ is the only S-wave rearrangement channel for the lowest $I(J^P)=\frac12(\frac32^-)$ state. There are four $I(J^P)=\frac32(\frac12^-)$ pentaquarks. The highest and the second highest states share the same main channels, with partial width ratios being $\Gamma_{\Sigma_c^*D^*}:\Gamma_{\Sigma_cD^*}:\Gamma_{\Sigma_cD}:\Gamma_{\Xi_{cc}^*\rho}:\Gamma_{\Xi_{cc}\rho}=29.0:11.0:1.0:114.0:9.0$ and $46.0:34.0:1.0:1.0:30.0$, respectively. The second lowest state mainly decays into $\Sigma_cD$, $\Xi_{cc}\rho$, and $\Xi_{cc}\pi$ while the lowest has the unique S-wave rearrangement channel $\Xi_{cc}\pi$. For the $I=\frac12$ partner states, the second lowest state mainly decays into $\Sigma_cD$, $\Xi_{cc}\omega$, $\Xi_{cc}\rho$, and $\Xi_{cc}\pi$, but the lowest state has only one S-wave rearrangement decay channel $\Xi_{cc}\pi$. The six major decay channels for the second highest state are $\Sigma_c^*D^*$, $\Sigma_cD^*$, $\Sigma_cD$, $\Xi_{cc}\rho$, $\Xi_{cc}^*\omega$, and $\Xi_{cc}\omega$, while the highest state has an additional dominant channel $\Xi_{cc}^*\rho$.

There are eight $I_{nn}=0,I=\frac12$ $ccnn\bar{n}$ pentaquarks, spanning from 3714.2 to 4583.8 MeV, the highest state is the only one with $J^P=\frac52^-$ and the lowest with $J^P=\frac12^-$. The $\frac52^-$ state has two dominant decay channels with $\Gamma_{\Xi_{cc}^*\rho}:\Gamma_{\Xi_{cc}^*\omega}=3.1$. The masses of the three $J^P=\frac32^-$ states range from 3813.5 to 4567.8 MeV. The highest primarily decays into $\Lambda_cD^*$, $\Xi_{cc}^*\rho$, $\Xi_{cc}^*\pi$, $\Xi_{cc}\rho$, and $\Xi_{cc}\omega$, while the intermediate state has similar dominant channels except for replacing $\Xi_{cc}^*\rho$ with $\Xi_{cc}^*\eta$. The lowest state seems to be stable or very narrow because all the considered $S$-wave rearrangement decay channels are kinetically forbidden. In the remaining four $J^P=\frac12^-$ states, the lowest is stable and the other three higher states exhibit different decay properties. The highest state has main channels $\Lambda_cD^*$, $\Lambda_cD$, $\Xi_{cc}^*\rho$, $\Xi_{cc}\rho$, and $\Xi_{cc}\omega$. For the second highest state, the channel $\Xi_{cc}^*\rho$ is forbidden and two additional channels, $\Xi_{cc}\pi$ and $\Xi_{cc}\eta$, should be taken into consideration. The third highest $J^P=\frac12^-$ state can decay into $\Lambda_cD$, $\Xi_{cc}\pi$, and $\Xi_{cc}\eta$.

For the study of the lowest $I_{nn}=0,I=\frac12$ $ccnn\bar{n}$ pentaquark in the compact picture, one can find several works in the literature. In a QCD sum rule calculation with a diquark-diquark-antiquark type current \cite{Wang:2018lhz}, a mass around 4.21 GeV was obtained. From a quark potential model calculation \cite{Giannuzzi:2019esi} in the diquark-diquark-antiquark configuration, a mass around 4.54 GeV was acquired. In a diquark-triquark configuration, the authors of Ref. \cite{Xing:2021yid} got a pentaquark mass around 3.8 GeV. Our estimated mass (3714 MeV) is the lowest one, even lower than the $\Xi_{cc}\pi$ threshold. Note that the investigation in the chiral effective theory \cite{Guo:2017vcf} has predicted a resonance around the $\Xi_{cc}\pi$ threshold. Some near-threshold structure seems to exist according to these results.
	
\subsection{Spectra and decay widths of $ccnn\bar{s}$ states}
	
\begin{table}[htbp]\caption{Numerical results for the $ccnn\bar{s}$ states. The CMI matrix $\langle H_{CMI}\rangle$, its eigenvalues, and pentaquark masses using Eq. \eqref{mass3detail} are listed in the second, third, and fifth columns, respectively, in units of MeV. The estimated masses using Eq. \eqref{mass2} with two different baryon-meson thresholds are given in the sixth and seven columns and those using Eq. \eqref{mass1} are shown in the last column. The corresponding base vectors of $\langle H_{CMI}\rangle$ can be found in table \ref{basis}.}\label{massofccnns}\resizebox{\linewidth}{!}{
  \begin{tabular}{cccccccc}\hline\hline
  	\multicolumn{8}{c}{$ccnn\bar{s}$ states with $I=1$}\\\hline
  	$J^P$&$\langle H_{CMI}\rangle$&Eigenvalue&Eigenvector&Mass&$\Sigma_cD_s$&$\Xi_{cc}K$&Upper limits\\\hline
  	$\frac52^-$&$\left(\begin{array}{c}136.0\end{array}\right)$&$\left(\begin{array}{c}136.0\end{array}\right)$&$\left[\begin{array}{c}\{1.00\}\end{array}\right]$&$\left(\begin{array}{c}4609.2\end{array}\right)$&$\left(\begin{array}{c}4658.7\end{array}\right)$&$\left(\begin{array}{c}4586.5\end{array}\right)$&$\left(\begin{array}{c}4850.2\end{array}\right)$\\$\frac32^-$&$\left(\begin{array}{cccc}-33.3&35.8&167.3&59.9\\35.8&80.8&-42.8&-5.2\\167.3&-42.8&73.2&-16.0\\59.9&-5.2&-16.0&36.9\end{array}\right)$&$\left(\begin{array}{c}200.2\\93.1\\46.9\\-182.6\end{array}\right)$&$\left[\begin{array}{cccc}\{-0.58,0.11,-0.79,-0.14\}\\\{-0.25,-0.94,0.09,-0.20\}\\\{0.07,-0.24,-0.25,0.93\}\\\{0.77,-0.20,-0.55,-0.25\}\end{array}\right]$&$\left(\begin{array}{c}4673.4\\4566.3\\4520.1\\4290.6\end{array}\right)$&$\left(\begin{array}{c}4722.9\\4615.8\\4569.6\\4340.1\end{array}\right)$&$\left(\begin{array}{c}4650.8\\4543.6\\4497.4\\4268.0\end{array}\right)$&$\left(\begin{array}{c}4914.4\\4807.3\\4761.1\\4531.6\end{array}\right)$\\$\frac12^-$&$\left(\begin{array}{cccc}36.3&45.3&-105.8&-37.9\\45.3&-20.8&181.6&-85.6\\-105.8&181.6&100.0&-16.0\\-37.9&-85.6&-16.0&111.7\end{array}\right)$&$\left(\begin{array}{c}265.6\\155.6\\5.9\\-199.9\end{array}\right)$&$\left[\begin{array}{cccc}\{-0.18,0.55,0.75,-0.34\}\\\{-0.58,-0.18,0.32,0.73\}\\\{-0.70,-0.38,-0.16,-0.58\}\\\{0.37,-0.73,0.56,-0.13\}\end{array}\right]$&$\left(\begin{array}{c}4738.8\\4628.8\\4479.1\\4273.3\end{array}\right)$&$\left(\begin{array}{c}4788.3\\4678.3\\4528.6\\4322.8\end{array}\right)$&$\left(\begin{array}{c}4716.1\\4606.1\\4456.4\\4250.7\end{array}\right)$&$\left(\begin{array}{c}4979.8\\4869.8\\4720.1\\4514.3\end{array}\right)$\\\hline
  	
  	\multicolumn{8}{c}{$ccnn\bar{s}$ states with $I=0$}\\\hline
  	$J^P$&$\langle H_{CMI}\rangle$&Eigenvalue&Eigenvector&Mass&$\Lambda_cD_s$&$\Xi_{cc}K$&Upper limits\\\hline$\frac52^-$&$\left(\begin{array}{c}126.4\end{array}\right)$&$\left(\begin{array}{c}126.4\end{array}\right)$&$\left[\begin{array}{c}\{1.00\}\end{array}\right]$&$\left(\begin{array}{c}4599.6\end{array}\right)$&$\left(\begin{array}{c}4635.7\end{array}\right)$&$\left(\begin{array}{c}4576.9\end{array}\right)$&$\left(\begin{array}{c}4840.6\end{array}\right)$\\$\frac32^-$&$\left(\begin{array}{ccc}-162.9&149.4&167.3\\149.4&15.2&-42.8\\167.3&-42.8&-120.8\end{array}\right)$&$\left(\begin{array}{c}113.2\\-17.7\\-364.0\end{array}\right)$&$\left[\begin{array}{ccc}\{0.58,0.77,0.28\}\\\{-0.34,0.54,-0.77\}\\\{0.74,-0.36,-0.57\}\end{array}\right]$&$\left(\begin{array}{c}4586.4\\4455.5\\4109.2\end{array}\right)$&$\left(\begin{array}{c}4622.5\\4491.6\\4145.3\end{array}\right)$&$\left(\begin{array}{c}4563.7\\4432.8\\4086.5\end{array}\right)$&$\left(\begin{array}{c}4827.4\\4696.5\\4350.2\end{array}\right)$\\$\frac12^-$&$\left(\begin{array}{cccc}-69.3&27.7&188.9&-105.8\\27.7&-134.4&0.0&65.6\\188.9&0.0&-158.4&181.6\\-105.8&65.6&181.6&-174.4\end{array}\right)$&$\left(\begin{array}{c}91.1\\-16.9\\-142.3\\-468.5\end{array}\right)$&$\left[\begin{array}{cccc}\{-0.67,-0.15,-0.68,-0.24\}\\\{-0.55,0.29,0.22,0.75\}\\\{-0.14,-0.93,0.29,0.17\}\\\{0.47,-0.16,-0.63,0.59\}\end{array}\right]$&$\left(\begin{array}{c}4564.3\\4456.3\\4330.9\\4004.7\end{array}\right)$&$\left(\begin{array}{c}4600.4\\4492.4\\4367.0\\4040.8\end{array}\right)$&$\left(\begin{array}{c}4541.6\\4433.6\\4308.3\\3982.1\end{array}\right)$&$\left(\begin{array}{c}4805.3\\4697.3\\4571.9\\4245.7\end{array}\right)$\\\hline\hline
\end{tabular}}
\end{table}

\begin{table}[htbp]\caption{Rearrangement decays for the $ccnn\bar{s}$ states. The numbers in the parentheses are $(100|\mathcal{M}|^2/{\cal C}^2,\ \Gamma_i)$. The pentaquark masses, their partial widths ($\Gamma_i$'s) and total width ($\Gamma_{sum}$) are given in units of MeV. The symbol `-' means that the corresponding channel is kinematically forbidden.}\label{decayofccnns}\resizebox{\linewidth}{!}{
\begin{tabular}{ccccccccc}\hline\hline
	$J^P$&Mass&\multicolumn{6}{c}{Channels}&$\Gamma_{sum}$\\\hline
\multicolumn{9}{c}{$ccnn\bar{s}$ states with $I=1$}\\\hline
&&$\Sigma_c^*D_s^*$&$\Xi_{cc}^*K^*$&&&&&\\
$\frac52^-$&$\left[\begin{array}{c}4609.2\end{array}\right]$&$\left[\begin{array}{c}(33.3,-)\end{array}\right]$&$\left[\begin{array}{c}(33.3,5.4)\end{array}\right]$&&&&&$\left[\begin{array}{c}5.4\end{array}\right]$\\
&&$\Sigma_c^*D_s^*$&$\Sigma_cD_s^*$&$\Sigma_c^*D_s$&$\Xi_{cc}^*K^*$&$\Xi_{cc}^*K$&$\Xi_{cc}K^*$&\\
$\frac32^-$&$\left[\begin{array}{c}4673.4\\4566.3\\4520.1\\4290.6\end{array}\right]$&$\left[\begin{array}{c}(9.7,2.4)\\(0.3,-)\\(33.7,-)\\(3.6,-)\end{array}\right]$&$\left[\begin{array}{c}(0.0,0.0)\\(29.2,0.8)\\(10.6,-)\\(4.6,-)\end{array}\right]$&$\left[\begin{array}{c}(3.5,1.8)\\(5.0,1.8)\\(16.1,3.6)\\(17.2,-)\end{array}\right]$&$\left[\begin{array}{c}(35.2,10.2)\\(8.5,-)\\(3.0,-)\\(0.5,-)\end{array}\right]$&$\left[\begin{array}{c}(0.1,0.0)\\(0.2,0.2)\\(3.3,1.8)\\(38.0,11.4)\end{array}\right]$&$\left[\begin{array}{c}(18.2,7.0)\\(23.5,5.2)\\(0.0,0.0)\\(2.8,-)\end{array}\right]$&$\left[\begin{array}{c}21.4\\8.0\\5.4\\11.4\end{array}\right]$\\&&$\Sigma_c^*D_s^*$&$\Sigma_cD_s^*$&$\Sigma_cD_s$&$\Xi_{cc}^*K^*$&$\Xi_{cc}K^*$&$\Xi_{cc}K$&\\$\frac12^-$&$\left[\begin{array}{c}4738.8\\4628.8\\4479.1\\4273.3\end{array}\right]$&$\left[\begin{array}{c}(16.5,6.4)\\(29.7,-)\\(6.8,-)\\(2.5,-)\end{array}\right]$&$\left[\begin{array}{c}(2.0,1.0)\\(17.7,5.4)\\(1.5,-)\\(15.0,-)\end{array}\right]$&$\left[\begin{array}{c}(0.1,0.0)\\(0.3,0.2)\\(35.4,10.8)\\(5.8,-)\end{array}\right]$&$\left[\begin{array}{c}(46.1,17.4)\\(2.3,0.4)\\(5.3,-)\\(1.8,-)\end{array}\right]$&$\left[\begin{array}{c}(1.9,0.8)\\(16.5,5.4)\\(16.3,-)\\(1.4,-)\end{array}\right]$&$\left[\begin{array}{c}(0.1,0.0)\\(0.1,0.0)\\(1.2,0.6)\\(40.2,14.8)\end{array}\right]$&$\left[\begin{array}{c}25.6\\11.4\\11.4\\14.8\end{array}\right]$\\\hline
\multicolumn{9}{c}{$ccnn\bar{s}$ states with $I=0$}\\\hline
&&&$\Xi_{cc}^*K^*$&&&&&\\
$\frac52^-$&$\left[\begin{array}{c}4599.6\end{array}\right]$&&$\left[\begin{array}{c}(66.7,8.8)\end{array}\right]$&&&&&$\left[\begin{array}{c}8.8\end{array}\right]$\\
&&&$\Lambda_cD_s^*$&&$\Xi_{cc}^*K^*$&$\Xi_{cc}^*K$&$\Xi_{cc}K^*$&\\
$\frac32^-$&$\left[\begin{array}{c}4586.4\\4455.5\\4109.2\end{array}\right]$&&$\left[\begin{array}{c}(2.5,1.4)\\(19.9,6.2)\\(10.9,-)\end{array}\right]$&&$\left[\begin{array}{c}(46.0,3.2)\\(6.7,-)\\(0.1,-)\end{array}\right]$&$\left[\begin{array}{c}(0.6,0.4)\\(2.3,1.0)\\(55.5,-)\end{array}\right]$&$\left[\begin{array}{c}(17.5,4.6)\\(37.8,-)\\(0.2,-)\end{array}\right]$&$\left[\begin{array}{c}9.6\\7.2\\-\end{array}\right]$\\
&&&$\Lambda_cD_s^*$&$\Lambda_cD_s$&$\Xi_{cc}^*K^*$&$\Xi_{cc}K^*$&$\Xi_{cc}K$&\\
$\frac12^-$&$\left[\begin{array}{c}4564.3\\4456.3\\4330.9\\4004.7\end{array}\right]$&&$\left[\begin{array}{c}(3.1,1.6)\\(17.7,5.4)\\(37.2,-)\\(0.4,-)\end{array}\right]$&$\left[\begin{array}{c}(0.3,0.2)\\(6.6,3.8)\\(21.7,8.0)\\(13.0,-)\end{array}\right]$&$\left[\begin{array}{c}(18.7,-)\\(21.0,-)\\(4.6,-)\\(0.0,-)\end{array}\right]$&$\left[\begin{array}{c}(44.3,9.6)\\(19.4,-)\\(0.1,-)\\(0.0,-)\end{array}\right]$&$\left[\begin{array}{c}(0.3,0.2)\\(1.9,1.0)\\(3.0,1.2)\\(53.2,-)\end{array}\right]$&$\left[\begin{array}{c}11.6\\10.2\\9.2\\-\end{array}\right]$\\\hline\hline\end{tabular}}\end{table}

\iffalse
\begin{table}[htbp]
	\caption{Stability of the doubly charmed pentaquarks in various studies. The meanings of "S" and "US" are "stable" and "unstable", respectively.}
	\resizebox{0.4\linewidth}{!}
	{\begin{tabular}{c|c|c}\hline 
Reference& $ccnn\bar{n}$ ($I_{nn}=0,\ IJ^P=\frac12\frac12^-$)& $ccnn\bar{s}$ ($IJ^P=0\frac12^-$)\\\hline 
This work&S &S\\ 
\cite{Park:2018oib}&&S \\ 
\cite{Xing:2021yid}&&S \\
\cite{Wang:2018lhz}&US&\\ 
\cite{Guo:2017vcf}&US&S \\ 
\cite{Park:2023ygm}&&US \\\hline
\end{tabular}}
\end{table}
\fi

Different from the $ccnn\bar{n}$ case, exotic structures, if observed, in the $ccnn\bar{s}$ system cannot be understood within the conventional baryon configuration. We show the mass-related results and decay widths of $ccnn\bar{s}$ states in tables \ref{massofccnns} and \ref{decayofccnns}, respectively. The mass spectra are drawn in Fig. \ref{figures} (c) and (d). It is evident that our estimated masses are lower than the upper limits and the values calculated using the $\Sigma_cD_s$($\Lambda_cD_s$) threshold, but higher than those obtained with the $\Xi_{cc}K$ threshold.

There are nine $I=1\ ccnn\bar{s}$ states ranging from 4273.3 to 4738.8 MeV. Both the highest and lowest states have $J^P=\frac12^-$. The $\frac52^-$ pentaquark located around 4609 MeV can decay into $\Xi_{cc}^*K^*$ through $S$-wave quark rearrangement. Four $\frac32^-$ pentaquarks are located from 4290.6 to 4673.4 MeV. The highest state dominantly decays into $\Sigma_c^*D_s^*$, $\Sigma_c^*D_s$, $\Xi_{cc}^*K^*$, and $\Xi_{cc}K^*$, while the partial widths for the remaining $S$-wave $\Sigma_cD_s^*$ and $\Xi_{cc}^*K$ channels are narrow. For the second highest state, the $\Sigma_c^*D_s^*$ and $\Xi_{cc}^*K^*$ channels are kinematically forbidden. The second lowest state mainly decays into $\Sigma_c^*D_s$ and $\Xi_{cc}^*K$ and the lowest state has only one $S$-wave rearrangement decay channel $\Xi_{cc}^*K$. Four $\frac12^-$ pentaquarks have the same possible decay channels as those of $\frac32^-$ states, except for replacing $\Sigma_c^*D_s$ and $\Xi_{cc}^*K$ with $\Sigma_cD_s$ and $\Xi_{cc}K$, respectively. The highest state has all possible channels open with very narrow partial widths for $\Sigma_cD_s$ and $\Xi_{cc}K$ channels. For the second highest state, the $\Sigma_c^*D_s^*$ channel is forbidden and the partial width for $\Xi_{cc}K$ is very narrow. The second lowest state has $\Sigma_cD_s$ and $\Xi_{cc}K$ open channels and the lowest state mainly decays into $\Xi_{cc}K$.

There are eight $I=0$ $ccnn\bar{s}$ states ranging from 4004.7 to 4599.6 MeV. The highest pentaquark is the only one with $J^P=\frac52^-$ that mainly decays into $\Xi_{cc}^*K^*$ through $S$-wave, and the lowest is the ground $\frac12^-$ state which is stable. The other three $\frac12^-$ states are located from 4330.9 to 4564.3 MeV. The second lowest $\frac12^-$ state has $S$-wave rearrangement channels $\Lambda_cD_s$ and $\Xi_{cc}K$, while the second highest state has an additional $S$-wave channel $\Lambda_cD_s^*$. For the highest $\frac12^-$ state, one more channel $\Xi_{cc}K^*$ should be considered. The three $\frac32^-$ states range from 4109.2 to 4586.4 MeV. The second highest state has two $S$-wave channels $\Lambda_cD_s^*$ and $\Xi_{cc}^*K$, while two additional channels, $\Xi_{cc}^*K^*$ and $\Xi_{cc}K^*$, should be taken into consideration for the highest pentaquark. The lowest state is close to the $\Xi_{cc}K$ threshold and is possibly narrow.

For the ground $ccnn\bar{s}$ pentaquark in the compact picture, a quark model calculation of Ref. \cite{Park:2018oib} gave a bound state about 135 MeV below the $\Xi_{cc}K$ threshold (4117.2 MeV) while a full model calculation of Ref. \cite{Park:2023ygm} obtained a resonance about 19 MeV above the $\Xi_{cc}K$ threshold. In the diquark-triquark configuration, the authors of Ref. \cite{Xing:2021yid} got a pentaquark mass around 4.1 GeV. Our mass (4005 MeV) is slightly larger than the lowest one. If our result is underestimated and one adds 100 MeV to the value, the mass is still below the $\Xi_{cc}K$ threshold. From the study of pseudoscalar meson scattering with doubly-charmed baryon \cite{Guo:2017vcf}, a pair of bound and virtual poles appear around the $\Xi_{cc}K$ threshold in the $I=0$ case. These studies indicate that some explicitly exotic structure(s) around the $\Xi_{cc}K$ threshold should exist.

\begin{table}[htbp]\caption{Numerical results for the $ccss\bar{n}$ and $ccss\bar{s}$ states. The CMI matrix $\langle H_{CMI}\rangle$, its eigenvalues, and pentaquark masses using Eq. \eqref{mass3detail} are listed in the second, third, and fifth columns, respectively, in units of MeV. The estimated masses using Eq. \eqref{mass2} with two different baryon-meson thresholds are given in the sixth and seven columns and those using Eq. \eqref{mass1} are shown in the last column. The corresponding base vectors of $\langle H_{CMI}\rangle$ can be found in table \ref{basis}.}\label{massofccssq}
	\resizebox{\linewidth}{!}{\begin{tabular}{cccccccc}\hline\hline
\multicolumn{8}{c}{$ccss\bar{n}$ states}\\\hline
$J^P$&$\langle H_{CMI}\rangle$&Eigenvalue&Eigenvector&Mass&$\Omega_cD$&$\Omega_{cc}\bar{K}$&Upper limits\\\hline$\frac52^-$&$\left(\begin{array}{c}102.7\end{array}\right)$&$\left(\begin{array}{c}102.7\end{array}\right)$&$\left[\begin{array}{c}\{1.00\}\end{array}\right]$&$\left(\begin{array}{c}4666.5\end{array}\right)$&$\left(\begin{array}{c}4801.4\end{array}\right)$&$\left(\begin{array}{c}4643.8\end{array}\right)$&$\left(\begin{array}{c}4997.5\end{array}\right)$\\$\frac32^-$&$\left(\begin{array}{cccc}-66.0&36.1&167.3&59.0\\36.1&46.0&-40.4&-8.0\\167.3&-40.4&22.5&-17.2\\59.0&-8.0&-17.2&3.1\end{array}\right)$&$\left(\begin{array}{c}155.7\\57.1\\13.7\\-220.9\end{array}\right)$&$\left[\begin{array}{cccc}\{-0.61,0.10,-0.77,-0.15\}\\\{-0.23,-0.96,0.09,-0.13\}\\\{0.07,-0.17,-0.26,0.95\}\\\{0.76,-0.20,-0.57,-0.25\}\end{array}\right]$&$\left(\begin{array}{c}4719.5\\4620.9\\4577.5\\4342.9\end{array}\right)$&$\left(\begin{array}{c}4854.5\\4755.9\\4712.4\\4477.9\end{array}\right)$&$\left(\begin{array}{c}4696.9\\4598.2\\4554.8\\4320.2\end{array}\right)$&$\left(\begin{array}{c}5050.5\\4951.9\\4908.5\\4673.9\end{array}\right)$\\$\frac12^-$&$\left(\begin{array}{cccc}0.8&45.6&-105.8&-37.3\\45.6&-55.2&184.0&-87.2\\-105.8&184.0&48.9&-17.2\\-37.3&-87.2&-17.2&77.9\end{array}\right)$&$\left(\begin{array}{c}225.4\\118.4\\-29.2\\-242.2\end{array}\right)$&$\left[\begin{array}{cccc}\{0.16,-0.56,-0.72,0.37\}\\\{-0.59,-0.14,0.35,0.71\}\\\{-0.70,-0.39,-0.16,-0.58\}\\\{0.37,-0.72,0.58,-0.12\}\end{array}\right]$&$\left(\begin{array}{c}4789.2\\4682.2\\4534.6\\4321.6\end{array}\right)$&$\left(\begin{array}{c}4924.2\\4817.1\\4669.5\\4456.6\end{array}\right)$&$\left(\begin{array}{c}4766.6\\4659.5\\4511.9\\4298.9\end{array}\right)$&$\left(\begin{array}{c}5120.2\\5013.2\\4865.6\\4652.6\end{array}\right)$\\\hline
\multicolumn{8}{c}{$ccss\bar{s}$ states}\\\hline$J^P$&$\langle H_{CMI}\rangle$&Eigenvalue&Eigenvector&Mass&$\Omega_cD_s$&$\Omega_{cc}\phi$&Upper limits\\\hline$\frac52^-$&$\left(\begin{array}{c}78.1\end{array}\right)$&$\left(\begin{array}{c}78.1\end{array}\right)$&$\left[\begin{array}{c}\{1.00\}\end{array}\right]$&$\left(\begin{array}{c}4732.5\end{array}\right)$&$\left(\begin{array}{c}4879.6\end{array}\right)$&$\left(\begin{array}{c}4793.8\end{array}\right)$&$\left(\begin{array}{c}5153.5\end{array}\right)$\\$\frac32^-$&$\left(\begin{array}{cccc}-29.2&8.0&84.1&59.9\\8.0&33.7&-3.2&-7.6\\84.1&-3.2&22.4&-17.2\\59.9&-7.6&-17.2&15.5\end{array}\right)$&$\left(\begin{array}{c}92.6\\37.9\\30.1\\-118.2\end{array}\right)$&$\left[\begin{array}{cccc}\{0.64,0.01,0.68,0.35\}\\\{0.03,-0.75,-0.32,0.58\}\\\{0.11,0.66,-0.42,0.61\}\\\{0.76,-0.07,-0.50,-0.41\}\end{array}\right]$&$\left(\begin{array}{c}4747.0\\4692.3\\4684.5\\4536.2\end{array}\right)$&$\left(\begin{array}{c}4894.1\\4839.4\\4831.5\\4683.2\end{array}\right)$&$\left(\begin{array}{c}4808.3\\4753.6\\4745.8\\4597.5\end{array}\right)$&$\left(\begin{array}{c}5168.0\\5113.3\\5105.5\\4957.2\end{array}\right)$\\$\frac12^-$&$\left(\begin{array}{cccc}0.8&10.2&-53.2&-37.9\\10.2&-30.7&109.6&-88.0\\-53.2&109.6&49.2&-17.2\\-37.9&-88.0&-17.2&53.1\end{array}\right)$&$\left(\begin{array}{c}167.8\\83.2\\-44.5\\-134.0\end{array}\right)$&$\left[\begin{array}{cccc}\{0.05,-0.58,-0.63,0.52\}\\\{-0.61,-0.04,0.50,0.61\}\\\{0.77,0.18,0.32,0.52\}\\\{0.17,-0.80,0.50,-0.29\}\end{array}\right]$&$\left(\begin{array}{c}4822.2\\4737.6\\4609.9\\4520.4\end{array}\right)$&$\left(\begin{array}{c}4969.2\\4884.7\\4756.9\\4667.4\end{array}\right)$&$\left(\begin{array}{c}4883.5\\4798.9\\4671.2\\4581.7\end{array}\right)$&$\left(\begin{array}{c}5243.2\\5158.6\\5030.9\\4941.4\end{array}\right)$\\\hline\hline\end{tabular}}\end{table}

\begin{table}[htbp]\caption{Rearrangement decays for the $ccss\bar{n}$ and $ccss\bar{s}$ states. The numbers in the parentheses are $(100|\mathcal{M}|^2/{\cal C}^2,\ \Gamma_i)$. The pentaquark masses, their partial widths ($\Gamma_i$'s) and total width ($\Gamma_{sum}$) are given in units of MeV. The symbol `-' means that the corresponding channel is kinematically forbidden.}\label{decayofccssq}\resizebox{\linewidth}{!}{
\begin{tabular}{cccccccccc}\hline\hline$J^P$&Mass&\multicolumn{7}{c}{Channels}&$\Gamma_{sum}$\\\hline\multicolumn{10}{c}{$ccss\bar{n}$ states}\\\hline&&$\Omega_c^*D^*$&$\Omega_{cc}^*\bar{K}^*$&&&&&&\\
$\frac52^-$&$\left[\begin{array}{c}4666.5\end{array}\right]$&$\left[\begin{array}{c}(33.3,-)\end{array}\right]$&$\left[\begin{array}{c}(33.3,-)\end{array}\right]$&&&&&&$\left[\begin{array}{c}-\end{array}\right]$
\\
&&$\Omega_c^*D^*$&$\Omega_cD^*$&$\Omega_c^*D$&$\Omega_{cc}^*\bar{K}^*$&$\Omega_{cc}\bar{K}^*$&$\Omega_{cc}^*\bar{K}$&&\\
$\frac32^-$&$\left[\begin{array}{c}4719.5\\4620.9\\4577.5\\4342.9\end{array}\right]$&$\left[\begin{array}{c}(10.3,-)\\(1.0,-)\\(32.4,-)\\(3.5,-)\end{array}\right]$&$\left[\begin{array}{c}(0.1,0.0)\\(26.6,-)\\(13.3,-)\\(4.4,-)\end{array}\right]$&$\left[\begin{array}{c}(3.8,1.4)\\(6.0,-)\\(15.2,-)\\(16.6,-)\end{array}\right]$&$\left[\begin{array}{c}(35.7,7.2)\\(8.3,-)\\(2.6,-)\\(0.7,-)\end{array}\right]$&$\left[\begin{array}{c}(16.8,5.4)\\(24.4,3.0)\\(0.2,-)\\(3.1,-)\end{array}\right]$&$\left[\begin{array}{c}(0.0,0.0)\\(0.3,0.2)\\(3.1,1.4)\\(38.3,8.8)\end{array}\right]$&&$\left[\begin{array}{c}14.0\\3.2\\1.4\\8.8\end{array}\right]$\\&&$\Omega_c^*D^*$&$\Omega_cD^*$&$\Omega_cD$&$\Omega_{cc}^*\bar{K}^*$&$\Omega_{cc}\bar{K}^*$&$\Omega_{cc}\bar{K}$&&\\$\frac12^-$&$\left[\begin{array}{c}4789.2\\4682.2\\4534.6\\4321.6\end{array}\right]$&$\left[\begin{array}{c}(19.1,2.6)\\(27.2,-)\\(6.8,-)\\(2.5,-)\end{array}\right]$&$\left[\begin{array}{c}(1.6,0.6)\\(18.6,-)\\(1.4,-)\\(14.6,-)\end{array}\right]$&$\left[\begin{array}{c}(0.1,0.0)\\(0.4,0.2)\\(35.7,-)\\(5.6,-)\end{array}\right]$&$\left[\begin{array}{c}(44.5,14.0)\\(3.4,0.4)\\(5.5,-)\\(2.2,-)\end{array}\right]$&$\left[\begin{array}{c}(1.4,0.6)\\(17.1,4.6)\\(16.2,-)\\(1.5,-)\end{array}\right]$&$\left[\begin{array}{c}(0.0,0.0)\\(0.1,0.0)\\(1.2,0.6)\\(40.4,12.4)\end{array}\right]$&&$\left[\begin{array}{c}17.8\\5.2\\0.6\\12.4\end{array}\right]$\\\hline\multicolumn{10}{c}{$ccss\bar{s}$ states}\\\hline&&$\Omega_c^*D_s^*$&$\Omega_{cc}^*\phi$&&&&&&\\$\frac52^-$&$\left[\begin{array}{c}4732.5\end{array}\right]$&$\left[\begin{array}{c}(33.3,-)\end{array}\right]$&$\left[\begin{array}{c}(33.3,-)\end{array}\right]$&&&&&&$\left[\begin{array}{c}-\end{array}\right]$\\&&$\Omega_c^*D_s^*$&$\Omega_cD_s^*$&$\Omega_c^*D_s$&$\Omega_{cc}^*\phi$&$\Omega_{cc}\phi$&$\Omega_{cc}^*\eta$&$\Omega_{cc}^*\eta'$&\\$\frac32^-$&$\left[\begin{array}{c}4747.0\\4692.3\\4684.5\\4536.2\end{array}\right]$&$\left[\begin{array}{c}(16.6,-)\\(26.5,-)\\(3.6,-)\\(0.5,-)\end{array}\right]$&$\left[\begin{array}{c}(2.8,-)\\(0.6,-)\\(36.0,-)\\(5.1,-)\end{array}\right]$&$\left[\begin{array}{c}(2.3,0.4)\\(14.1,-)\\(0.4,-)\\(24.8,-)\end{array}\right]$&$\left[\begin{array}{c}(34.7,-)\\(12.5,-)\\(0.0,-)\\(0.0,-)\end{array}\right]$&$\left[\begin{array}{c}(10.2,1.4)\\(4.2,-)\\(25.9,-)\\(4.2,-)\end{array}\right]$&$\left[\begin{array}{c}(0.0,0.0)\\(4.2,2.2)\\(0.4,0.2)\\(15.2,6.0)\end{array}\right]$&$\left[\begin{array}{c}(0.0,0.0)\\(4.6,-)\\(0.4,-)\\(16.9,-)\end{array}\right]$&$\left[\begin{array}{c}1.8\\2.2\\0.2\\6.0\end{array}\right]$\\&&$\Omega_c^*D_s^*$&$\Omega_cD_s^*$&$\Omega_cD_s$&$\Omega_{cc}^*\phi$&$\Omega_{cc}\phi$&$\Omega_{cc}\eta$&$\Omega_{cc}\eta'$&\\$\frac12^-$&$\left[\begin{array}{c}4822.2\\4737.6\\4609.9\\4520.4\end{array}\right]$&$\left[\begin{array}{c}(29.0,-)\\(17.9,-)\\(8.2,-)\\(0.5,-)\end{array}\right]$&$\left[\begin{array}{c}(0.1,0.0)\\(19.3,-)\\(4.9,-)\\(11.8,-)\end{array}\right]$&$\left[\begin{array}{c}(0.0,0.0)\\(0.3,0.0)\\(25.8,-)\\(15.6,-)\end{array}\right]$&$\left[\begin{array}{c}(37.3,5.6)\\(10.2,-)\\(7.5,-)\\(0.5,-)\end{array}\right]$&$\left[\begin{array}{c}(0.1,0.0)\\(18.8,1.8)\\(12.4,-)\\(4.8,-)\end{array}\right]$&$\left[\begin{array}{c}(0.0,0.0)\\(0.1,0.0)\\(3.7,2.0)\\(15.9,7.2)\end{array}\right]$&$\left[\begin{array}{c}(0.0,0.0)\\(0.1,0.0)\\(4.2,-)\\(17.7,-)\end{array}\right]$&$\left[\begin{array}{c}5.6\\1.8\\2.0\\7.2\end{array}\right]$\\\hline\hline
\end{tabular}}
\end{table}

\subsection{Spectra and decay widths of $ccss\bar{n}$ and $ccss\bar{s}$ states}

The numerical results for the $ccss\bar{n}$ and $ccss\bar{s}$ systems are collected in tables \ref{massofccssq} and \ref{decayofccssq}. We show their spectra in Fig. \ref{figures} (e) and (f), respectively. Similar to the $ccnn\bar{s}$ case, one would not understand the observed exotic structures in the $ccss\bar{n}$ system with the conventional baryon configuration.

The obtained masses in the $ccss\bar{n}$ case fall below the upper limits obtained with Eq. \eqref{mass1}. They lie between the estimated values based on the $\Omega_cD$ threshold and the values using $\Omega_{cc}\bar{K}$, and are closer to the latter ones. There are nine $ccss\bar{n}$ pentaquarks ranging from 4321.6 to 4789.2 MeV, with both the highest and lowest states having $J^P=\frac12^-$. All of them have strong decay modes. The $J^P=\frac52^-$ state slightly below the $\Omega_{cc}^*K^*$ threshold possibly has a narrow width. The four $\frac32^-$ states range from 4342.9 to 4719.5 MeV. The highest one has three dominant channels, $\Omega_c^*D$, $\Omega_{cc}^*\bar{K}^*$, and $\Omega_{cc}\bar{K}^*$. The second highest mainly decays into $\Omega_{cc}^*\bar{K}$ and $\Omega_{cc}\bar{K}^*$. The two lower states both dominantly decay into $\Omega_{cc}^*\bar{K}$, but the width of the lowest state seems to be broader. For the four $\frac12^-$ states, the highest one has four main decay channels, $\Omega_c^*D^*$, $\Omega_cD^*$, $\Omega_{cc}^*\bar{K}^*$, and $\Omega_{cc}\bar{K}^*$. The second highest has dominant channels $\Omega_cD$, $\Omega_{cc}^*\bar{K}^*$, and $\Omega_{cc}\bar{K}^*$. The $\Omega_{cc}\bar{K}$ is the main channel for the lower two state decays, but the lowest pentaquark has much larger width. 

The $ccss\bar{s}$ case has some similar features with the $ccss\bar{n}$ case. Our masses, which are much smaller than the theoretical upper limits, are between the estimated values using the $\Omega_cD_s$ threshold and those using $\Omega_{cc}\phi$. They are closer to the latter values. All the nine pentaquarks which are distributed between 4520.4 and 4822.2 MeV have strong decay channels. Both the highest and lowest states possess the spin-parity of $J^P=\frac12^-$. The near-threshold $\frac52^-$ state at 4732.5 MeV is probably narrow. For the four $\frac32^-$ states ranging from 4536.2 to 4747.0 MeV, the highest mainly decays into $\Omega_c^*D_s$ and $\Omega_{cc}\phi$ and the other three into $\Omega_{cc}^*\eta$. The width of the lowest (second lowest) pentaquark is larger (smaller) than those of the other three states. Of the four $\frac12^-$ pentaquarks, the highest primarily decays into $\Omega_{cc}^*\phi$ while the second highest dominantly decays into $\Omega_{cc}\phi$. The lower two states mainly decay into $\Omega_{cc}\eta$, with the lighter being broader. Since the $ccss\bar{s}$ states apparently have the same quantum numbers as the excited $\Omega_{cc}$ baryons, mixing of different configurations cannot be excluded, which may induce changes of the spectrum.

\begin{table}[htbp]\caption{Numerical results for the $ccns\bar{n}$ and $ccns\bar{s}$ states. The CMI matrix $\langle H_{CMI}\rangle$, its eigenvalues, and pentaquark masses using Eq. \eqref{mass3detail} are listed in the second, third, and fifth columns, respectively, in units of MeV. The estimated masses using Eq. \eqref{mass2} with two different baryon-meson thresholds are given in the sixth and seven columns and those using Eq. \eqref{mass1} are shown in the last column. The corresponding base vectors of $\langle H_{CMI}\rangle$ can be found in table \ref{basis}.}\label{massofccnsq}
	\resizebox{\linewidth}{!}{
\begin{tabular}{cccccccc}\hline\hline
\multicolumn{8}{c}{$ccns\bar{n}$ states with $I=1,0$}\\\hline
$J^P$&$\langle H_{CMI}\rangle$&Eigenvalue&Eigenvector&Mass&$\Xi_cD$&$\Xi_{cc}\bar{K}$&Upper limits\\\hline
	$\frac52^-$&$\left(\begin{array}{cc}173.3&-32.5\\-32.5&134.3\end{array}\right)$&$\left(\begin{array}{c}191.7\\115.8\end{array}\right)$&$\left[\begin{array}{cc}\{-0.87,0.49\}\\\{-0.49,-0.87\}\end{array}\right]$&$\left(\begin{array}{c}4664.9\\4589.0\end{array}\right)$&$\left(\begin{array}{c}4730.5\\4654.6\end{array}\right)$&$\left(\begin{array}{c}4642.2\\4566.4\end{array}\right)$&$\left(\begin{array}{c}4905.9\\4830.0\end{array}\right)$\\$\frac32^-$&$\left(\begin{array}{ccccccc}-209.7&46.7&191.0&-35.4&-59.0&217.3&0.0\\46.7&-71.7&-35.4&52.8&217.3&-23.6&59.0\\191.0&-35.4&41.3&-15.0&25.0&-64.0&-0.8\\-35.4&52.8&-15.0&70.9&-64.0&10.0&-6.8\\-59.0&217.3&25.0&-64.0&48.1&0.0&-16.6\\217.3&-23.6&-64.0&10.0&0.0&-70.7&-0.8\\0.0&59.0&-0.8&-6.8&-16.6&-0.8&12.7\end{array}\right)$&$\left(\begin{array}{c}223.2\\169.0\\83.4\\42.7\\19.7\\-237.2\\-479.9\end{array}\right)$&$\left[\begin{array}{ccccccc}\{0.16,-0.57,0.11,0.09,-0.77,0.14,-0.10\}\\\{0.52,0.05,0.71,-0.36,0.17,0.26,0.01\}\\\{-0.26,-0.31,-0.08,-0.83,0.03,-0.33,-0.18\}\\\{-0.21,-0.04,0.58,0.34,0.00,-0.69,-0.15\}\\\{-0.05,0.04,0.07,-0.14,-0.21,-0.20,0.94\}\\\{0.25,0.70,-0.11,-0.20,-0.53,-0.26,-0.21\}\\\{-0.73,0.28,0.35,-0.10,-0.23,0.46,-0.04\}\end{array}\right]$&$\left(\begin{array}{c}4696.4\\4642.2\\4556.6\\4515.9\\4492.9\\4236.0\\3993.3\end{array}\right)$&$\left(\begin{array}{c}4762.0\\4707.8\\4622.2\\4581.5\\4558.5\\4301.6\\4058.9\end{array}\right)$&$\left(\begin{array}{c}4673.8\\4619.5\\4534.0\\4493.3\\4470.2\\4213.3\\3970.6\end{array}\right)$&$\left(\begin{array}{c}4937.4\\4883.2\\4797.6\\4756.9\\4733.9\\4477.0\\4234.3\end{array}\right)$\\$\frac12^-$&$\left(\begin{array}{cccccccc}-62.9&1.7&28.8&241.5&-44.8&37.3&-137.5&0.0\\1.7&18.7&1.5&-44.8&66.8&-137.5&14.9&-37.3\\28.8&1.5&-84.0&0.0&0.0&0.0&64.7&12.9\\241.5&-44.8&0.0&-188.5&32.5&-54.2&227.6&-0.8\\-44.8&66.8&0.0&32.5&-52.7&227.6&-21.7&-86.0\\37.3&-137.5&0.0&-54.2&227.6&74.5&0.0&-16.6\\-137.5&14.9&64.7&227.6&-21.7&0.0&-123.5&-0.8\\0.0&-37.3&12.9&-0.8&-86.0&-16.6&-0.8&109.9\end{array}\right)$&$\left(\begin{array}{c}285.9\\164.9\\132.8\\46.0\\-0.7\\-95.1\\-267.9\\-574.5\end{array}\right)$&$\left[\begin{array}{cccccccc}\{0.01,0.20,0.03,0.07,-0.54,-0.75,0.08,0.29\}\\\{0.38,-0.55,0.09,0.32,-0.18,0.25,0.08,0.59\}\\\{-0.51,-0.16,-0.13,-0.60,-0.09,0.11,-0.29,0.48\}\\\{0.56,0.01,-0.26,-0.16,-0.13,-0.05,-0.75,-0.14\}\\\{-0.21,-0.69,-0.11,-0.05,-0.39,-0.12,0.07,-0.54\}\\\{0.08,-0.03,0.94,-0.25,-0.12,0.02,-0.17,-0.11\}\\\{0.10,0.37,-0.10,-0.15,-0.67,0.55,0.24,-0.09\}\\\{0.48,-0.13,-0.09,-0.65,0.20,-0.18,0.50,0.02\}\end{array}\right]$&$\left(\begin{array}{c}4759.1\\4638.1\\4606.0\\4519.2\\4472.5\\4378.1\\4205.3\\3898.7\end{array}\right)$&$\left(\begin{array}{c}4824.7\\4703.7\\4671.6\\4584.8\\4538.1\\4443.7\\4270.9\\3964.3\end{array}\right)$&$\left(\begin{array}{c}4736.5\\4615.5\\4583.3\\4496.6\\4449.9\\4355.5\\4182.6\\3876.0\end{array}\right)$&$\left(\begin{array}{c}5000.1\\4879.1\\4847.0\\4760.2\\4713.5\\4619.1\\4446.3\\4139.7\end{array}\right)$\\\hline
\multicolumn{8}{c}{$ccns\bar{s}$ states}\\\hline
$J^P$&$\langle H_{CMI}\rangle$&Eigenvalue&Eigenvector&Mass&$\Xi_cD_s$&$\Xi_{cc}\phi$&Upper limits\\\hline
$\frac52^-$&$\left(\begin{array}{cc}104.8&-27.2\\-27.2&107.2\end{array}\right)$&$\left(\begin{array}{c}133.2\\78.8\end{array}\right)$&$\left[\begin{array}{cc}\{-0.69,0.72\}\\\{-0.72,-0.69\}\end{array}\right]$&$\left(\begin{array}{c}4697.0\\4642.6\end{array}\right)$&$\left(\begin{array}{c}4774.7\\4720.3\end{array}\right)$&$\left(\begin{array}{c}4758.3\\4703.9\end{array}\right)$&$\left(\begin{array}{c}5028.0\\4973.6\end{array}\right)$\\$\frac32^-$&$\left(\begin{array}{ccccccc}-107.0&38.6&114.7&-29.4&-49.0&125.7&0.0\\38.6&-31.1&-29.4&21.9&125.7&-19.6&59.9\\114.7&-29.4&7.1&-12.3&20.5&-23.0&-0.8\\-29.4&21.9&-12.3&57.4&-23.0&8.2&-6.4\\-49.0&125.7&20.5&-23.0&48.0&0.0&-16.6\\125.7&-19.6&-23.0&8.2&0.0&-70.4&-0.8\\0.0&59.9&-0.8&-6.4&-16.6&-0.8&26.3\end{array}\right)$&$\left(\begin{array}{c}149.7\\106.0\\56.0\\40.8\\-10.6\\-121.9\\-289.8\end{array}\right)$&$\left[\begin{array}{ccccccc}\{0.19,-0.58,0.14,-0.01,-0.74,0.15,-0.18\}\\\{-0.52,-0.12,-0.61,0.48,-0.20,-0.26,-0.08\}\\\{0.29,0.27,0.13,0.70,-0.18,0.26,0.49\}\\\{-0.09,0.04,-0.29,-0.50,-0.30,-0.09,0.75\}\\\{0.24,0.13,-0.62,-0.14,0.03,0.69,-0.21\}\\\{0.25,0.68,-0.07,-0.10,-0.46,-0.37,-0.33\}\\\{-0.69,0.33,0.35,-0.10,-0.25,0.46,-0.07\}\end{array}\right]$&$\left(\begin{array}{c}4713.5\\4669.8\\4619.8\\4604.6\\4553.2\\4441.9\\4274.0\end{array}\right)$&$\left(\begin{array}{c}4791.2\\4747.5\\4697.5\\4682.3\\4630.9\\4519.6\\4351.7\end{array}\right)$&$\left(\begin{array}{c}4774.8\\4731.1\\4681.1\\4665.9\\4614.5\\4503.2\\4335.3\end{array}\right)$&$\left(\begin{array}{c}5044.5\\5000.8\\4950.8\\4935.6\\4884.2\\4772.9\\4605.0\end{array}\right)$\\$\frac12^-$&$\left(\begin{array}{cccccccc}-62.9&1.7&28.8&145.1&-37.2&31.0&-79.5&0.0\\1.7&18.7&1.5&-37.2&27.7&-79.5&12.4&-37.9\\28.8&1.5&-84.0&0.0&0.0&0.0&65.6&10.7\\145.1&-37.2&0.0&-120.0&27.2&-45.3&145.6&-0.8\\-37.2&27.7&0.0&27.2&-25.6&145.6&-18.1&-86.8\\31.0&-79.5&0.0&-45.3&145.6&74.8&0.0&-16.6\\-79.5&12.4&65.6&145.6&-18.1&0.0&-124.0&-0.8\\0.0&-37.9&10.7&-0.8&-86.8&-16.6&-0.8&82.5\end{array}\right)$&$\left(\begin{array}{c}216.9\\124.3\\76.8\\-3.0\\-22.5\\-91.7\\-154.3\\-387.0\end{array}\right)$&$\left[\begin{array}{cccccccc}\{0.02,0.12,0.03,0.07,-0.56,-0.70,0.07,0.42\}\\\{-0.17,0.60,-0.05,-0.10,0.12,-0.40,0.01,-0.65\}\\\{0.53,-0.06,0.22,0.68,0.04,-0.07,0.35,-0.26\}\\\{0.59,-0.08,-0.35,-0.08,-0.19,-0.05,-0.66,-0.20\}\\\{0.32,0.73,0.14,0.01,0.24,0.22,-0.10,0.48\}\\\{-0.14,0.03,-0.83,0.36,0.31,-0.11,0.12,0.20\}\\\{0.01,0.27,-0.31,-0.03,-0.66,0.50,0.35,-0.15\}\\\{-0.46,0.12,0.16,0.61,-0.21,0.18,-0.54,-0.03\}\end{array}\right]$&$\left(\begin{array}{c}4780.7\\4688.1\\4640.6\\4560.8\\4541.3\\4472.1\\4409.5\\4176.8\end{array}\right)$&$\left(\begin{array}{c}4858.4\\4765.8\\4718.3\\4638.5\\4619.0\\4549.8\\4487.2\\4254.5\end{array}\right)$&$\left(\begin{array}{c}4842.0\\4749.4\\4701.9\\4622.1\\4602.6\\4533.4\\4470.8\\4238.1\end{array}\right)$&$\left(\begin{array}{c}5111.7\\5019.1\\4971.6\\4891.8\\4872.3\\4803.1\\4740.5\\4507.8\end{array}\right)$\\\hline\hline\end{tabular}}\end{table}

\begin{table}[htbp]\caption{Rearrangement decays for the $ccns\bar{n}$ and $ccns\bar{s}$ states. The numbers in the parentheses are $(100|\mathcal{M}|^2/{\cal C}^2,\ \Gamma_i)$. The pentaquark masses, their partial widths ($\Gamma_i$'s) and total width ($\Gamma_{sum}$) are given in units of MeV. The symbol `-' means that the corresponding channel is kinematically forbidden.}\label{decayofccnsq}\resizebox{\linewidth}{!}{
\begin{tabular}{ccccccccccccccc}\hline\hline
$J^P$&Mass&\multicolumn{12}{c}{Channels}&$\Gamma_{sum}$\\\hline
\multicolumn{15}{c}{$ccns\bar{n}$ states with $I=1$}\\\hline
&&$\Xi_c^*D^*$&$\Omega_{cc}^*\rho$&$\Xi_{cc}^*\bar{K}^*$&&&&&&&&&&\\
$\frac52^-$&$\left[\begin{array}{c}4664.9\\4589.0\end{array}\right]$&$\left[\begin{array}{c}(8.1,1.0)\\(25.2,-)\end{array}\right]$&$\left[\begin{array}{c}(99.0,15.2)\\(1.0,0.1)\end{array}\right]$&$\left[\begin{array}{c}(18.1,2.5)\\(81.9,3.6)\end{array}\right]$&&&&&&&&&&$\left[\begin{array}{c}18.7\\3.7\end{array}\right]$
\\
&&$\Xi_c^*D^*$&$\Xi_cD^*$&$\Xi_c'D^*$&$\Xi_c^*D$&&$\Omega_{cc}^*\rho$&$\Omega_{cc}\rho$&$\Omega_{cc}^*\pi$&&$\Xi_{cc}^*\bar{K}^*$&$\Xi_{cc}\bar{K}^*$&$\Xi_{cc}^*\bar{K}$&\\
$\frac32^-$&$\left[\begin{array}{c}4696.4\\4642.2\\4556.6\\4515.9\\4492.9\\4236.0\\3993.3\end{array}\right]$&$\left[\begin{array}{c}(7.9,2.0)\\(1.4,-)\\(0.0,-)\\(3.2,-)\\(30.0,-)\\(3.7,-)\\(1.1,-)\end{array}\right]$&$\left[\begin{array}{c}(0.7,0.4)\\(2.3,1.2)\\(3.7,1.2)\\(16.0,4.0)\\(1.3,0.2)\\(2.3,-)\\(7.0,-)\end{array}\right]$&$\left[\begin{array}{c}(0.0,0.0)\\(2.6,0.8)\\(22.7,-)\\(0.8,-)\\(14.2,-)\\(3.7,-)\\(0.4,-)\end{array}\right]$&$\left[\begin{array}{c}(4.0,2.0)\\(0.3,0.2)\\(5.6,1.4)\\(1.2,0.0)\\(15.8,-)\\(13.2,-)\\(1.5,-)\end{array}\right]$&&$\left[\begin{array}{c}(57.4,9.9)\\(36.6,5.0)\\(0.1,0.0)\\(5.0,-)\\(0.4,-)\\(0.5,-)\\(0.0,-)\end{array}\right]$&$\left[\begin{array}{c}(14.3,3.0)\\(40.1,7.4)\\(21.3,2.7)\\(21.5,1.8)\\(0.7,0.0)\\(1.9,-)\\(0.0,-)\end{array}\right]$&$\left[\begin{array}{c}(0.0,0.0)\\(0.1,0.0)\\(0.3,0.1)\\(0.0,0.0)\\(1.5,0.4)\\(9.3,1.8)\\(88.7,7.7)\end{array}\right]$&&$\left[\begin{array}{c}(17.0,2.8)\\(40.8,4.9)\\(35.4,-)\\(0.5,-)\\(4.2,-)\\(0.9,-)\\(1.2,-)\end{array}\right]$&$\left[\begin{array}{c}(19.0,3.9)\\(1.8,0.3)\\(12.1,1.2)\\(60.3,-)\\(2.0,-)\\(2.8,-)\\(2.0,-)\end{array}\right]$&$\left[\begin{array}{c}(0.4,0.1)\\(0.7,0.2)\\(0.1,0.0)\\(3.6,0.9)\\(1.9,0.5)\\(72.0,7.7)\\(21.4,-)\end{array}\right]$&$\left[\begin{array}{c}24.1\\20.0\\6.6\\6.7\\1.1\\9.5\\7.7\end{array}\right]$
\\
&&$\Xi_c^*D^*$&$\Xi_cD^*$&$\Xi_c'D^*$&$\Xi_c'D$&$\Xi_cD$&$\Omega_{cc}^*\rho$&$\Omega_{cc}\rho$&$\Omega_{cc}\pi$&&$\Xi_{cc}^*\bar{K}^*$&$\Xi_{cc}\bar{K}^*$&$\Xi_{cc}\bar{K}$&\\
$\frac12^-$&$\left[\begin{array}{c}4759.1\\4638.1\\4606.0\\4519.2\\4472.5\\4378.1\\4205.3\\3898.7\end{array}\right]$&$\left[\begin{array}{c}(14.1,5.2)\\(20.2,-)\\(12.5,-)\\(0.3,-)\\(5.3,-)\\(0.1,-)\\(2.7,-)\\(0.3,-)\end{array}\right]$&$\left[\begin{array}{c}(0.2,0.0)\\(0.7,0.4)\\(3.0,1.4)\\(16.1,4.2)\\(0.4,-)\\(37.3,-)\\(0.0,-)\\(0.6,-)\end{array}\right]$&$\left[\begin{array}{c}(2.6,1.2)\\(13.9,3.8)\\(3.0,0.6)\\(0.5,-)\\(1.5,-)\\(0.2,-)\\(13.3,-)\\(1.3,-)\end{array}\right]$&$\left[\begin{array}{c}(0.2,0.2)\\(0.0,0.0)\\(1.3,0.6)\\(1.2,0.4)\\(33.7,7.0)\\(1.0,-)\\(4.0,-)\\(0.3,-)\end{array}\right]$&$\left[\begin{array}{c}(0.1,0.0)\\(0.0,0.0)\\(0.9,0.6)\\(7.1,3.8)\\(0.7,0.4)\\(22.0,5.8)\\(2.6,-)\\(8.4,-)\end{array}\right]$&$\left[\begin{array}{c}(56.5,11.5)\\(0.6,0.1)\\(21.6,2.3)\\(12.3,-)\\(4.9,-)\\(3.1,-)\\(1.1,-)\\(0.0,-)\end{array}\right]$&$\left[\begin{array}{c}(1.8,0.4)\\(54.2,9.9)\\(14.6,2.4)\\(23.0,2.1)\\(5.4,-)\\(0.3,-)\\(0.7,-)\\(0.0,-)\end{array}\right]$&$\left[\begin{array}{c}(0.0,0.0)\\(0.1,0.0)\\(0.0,0.0)\\(0.5,0.1)\\(0.3,0.1)\\(0.6,0.2)\\(16.5,3.6)\\(82.0,5.9)\end{array}\right]$&&$\left[\begin{array}{c}(37.7,7.5)\\(8.8,1.0)\\(14.1,1.1)\\(25.4,-)\\(4.1,-)\\(4.9,-)\\(3.6,-)\\(1.5,-)\end{array}\right]$&$\left[\begin{array}{c}(3.0,0.7)\\(0.0,0.0)\\(40.8,6.0)\\(20.2,0.5)\\(34.0,-)\\(0.2,-)\\(1.4,-)\\(0.4,-)\end{array}\right]$&$\left[\begin{array}{c}(0.2,0.1)\\(0.0,0.0)\\(0.8,0.2)\\(1.9,0.5)\\(1.5,0.4)\\(3.0,0.7)\\(64.9,9.0)\\(27.7,-)\end{array}\right]$&$\left[\begin{array}{c}26.8\\15.2\\15.2\\11.6\\7.9\\6.7\\12.6\\5.9\end{array}\right]$\\\hline

\multicolumn{15}{c}{$ccns\bar{n}$ states with $I=0$}\\\hline
&&$\Xi_c^*D^*$&$\Omega_{cc}^*\omega$&$\Xi_{cc}^*\bar{K}^*$&&&&&&&&&&\\$\frac52^-$&$\left[\begin{array}{c}4664.9\\4589.0\end{array}\right]$&$\left[\begin{array}{c}(8.1,1.0)\\(25.2,-)\end{array}\right]$&$\left[\begin{array}{c}(99.0,14.7)\\(1.0,0.1)\end{array}\right]$&$\left[\begin{array}{c}(18.1,2.5)\\(81.9,3.6)\end{array}\right]$&&&&&&&&&&$\left[\begin{array}{c}18.2\\3.7\end{array}\right]$
\\
&&$\Xi_c^*D^*$&$\Xi_{c}D^*$&$\Xi_c'D^*$&$\Xi_c^*D$&&$\Omega_{cc}^*\omega$&$\Omega_{cc}\omega$&$\Omega_{cc}^*\eta$&$\Omega_{cc}^*\eta'$&$\Xi_{cc}^*\bar{K}^*$&$\Xi_{cc}\bar{K}^*$&$\Xi_{cc}^*\bar{K}$&\\
$\frac32^-$&$\left[\begin{array}{c}4696.4\\4642.2\\4556.6\\4515.9\\4492.9\\4236.0\\3993.3\end{array}\right]$&$\left[\begin{array}{c}(7.9,2.0)\\(1.4,-)\\(0.0,-)\\(3.2,-)\\(30.0,-)\\(3.7,-)\\(1.1,-)\end{array}\right]$&$\left[\begin{array}{c}(0.7,0.4)\\(2.3,1.2)\\(3.7,1.2)\\(16.0,4.0)\\(1.3,0.2)\\(2.3,-)\\(7.0,-)\end{array}\right]$&$\left[\begin{array}{c}(0.0,0.0)\\(2.6,0.8)\\(22.7,-)\\(0.8,-)\\(14.2,-)\\(3.7,-)\\(0.4,-)\end{array}\right]$&$\left[\begin{array}{c}(4.0,2.0)\\(0.3,0.2)\\(5.6,1.4)\\(1.2,0.0)\\(15.8,-)\\(13.2,-)\\(1.5,-)\end{array}\right]$&&$\left[\begin{array}{c}(57.4,9.7)\\(36.6,4.8)\\(0.1,-)\\(5.0,-)\\(0.4,-)\\(0.5,-)\\(0.0,-)\end{array}\right]$&$\left[\begin{array}{c}(14.3,3.0)\\(40.1,7.3)\\(21.3,2.6)\\(21.5,1.6)\\(0.7,0.0)\\(1.9,-)\\(0.0,-)\end{array}\right]$&$\left[\begin{array}{c}(0.0,0.0)\\(0.0,0.0)\\(0.2,0.0)\\(0.0,0.0)\\(0.8,0.1)\\(4.9,-)\\(46.8,-)\end{array}\right]$&$\left[\begin{array}{c}(0.0,-)\\(0.0,-)\\(0.2,-)\\(0.0,-)\\(0.7,-)\\(4.4,-)\\(41.9,-)\end{array}\right]$&$\left[\begin{array}{c}(17.0,2.8)\\(40.8,4.9)\\(35.4,-)\\(0.5,-)\\(4.2,-)\\(0.9,-)\\(1.2,-)\end{array}\right]$&$\left[\begin{array}{c}(19.0,3.9)\\(1.8,0.3)\\(12.1,1.2)\\(60.3,-)\\(2.0,-)\\(2.8,-)\\(2.0,-)\end{array}\right]$&$\left[\begin{array}{c}(0.4,0.1)\\(0.7,0.2)\\(0.1,0.0)\\(3.6,0.9)\\(1.9,0.5)\\(72.0,7.7)\\(21.4,-)\end{array}\right]$&$\left[\begin{array}{c}23.9\\19.7\\6.4\\6.5\\0.8\\7.7\\-\end{array}\right]$\\
&&$\Xi_c^*D^*$&$\Xi_cD^*$&$\Xi_c'D^*$&$\Xi_c'D$&$\Xi_cD$&$\Omega_{cc}^*\omega$&$\Omega_{cc}\omega$&$\Omega_{cc}\eta$&$\Omega_{cc}\eta'$&$\Xi_{cc}^*\bar{K}^*$&$\Xi_{cc}\bar{K}^*$&$\Xi_{cc}\bar{K}$&\\$\frac12^-$&$\left[\begin{array}{c}4759.1\\4638.1\\4606.0\\4519.2\\4472.5\\4378.1\\4205.3\\3898.7\end{array}\right]$&$\left[\begin{array}{c}(14.1,5.2)\\(20.2,-)\\(12.5,-)\\(0.3,-)\\(5.3,-)\\(0.1,-)\\(2.7,-)\\(0.3,-)\end{array}\right]$&$\left[\begin{array}{c}(0.2,0.0)\\(0.7,0.4)\\(3.0,1.4)\\(16.1,4.2)\\(0.4,-)\\(37.3,-)\\(0.0,-)\\(0.6,-)\end{array}\right]$&$\left[\begin{array}{c}(2.6,1.2)\\(13.9,3.8)\\(3.0,0.6)\\(0.5,-)\\(1.5,-)\\(0.2,-)\\(13.3,-)\\(1.3,-)\end{array}\right]$&$\left[\begin{array}{c}(0.2,0.2)\\(0.0,0.0)\\(1.3,0.6)\\(1.2,0.4)\\(33.7,7.0)\\(1.0,-)\\(4.0,-)\\(0.3,-)\end{array}\right]$&$\left[\begin{array}{c}(0.1,0.0)\\(0.0,0.0)\\(0.9,0.6)\\(7.1,3.8)\\(0.7,0.4)\\(22.0,5.8)\\(2.6,-)\\(8.4,-)\end{array}\right]$&$\left[\begin{array}{c}(56.5,11.3)\\(0.6,0.1)\\(21.6,2.1)\\(12.3,-)\\(4.9,-)\\(3.1,-)\\(1.1,-)\\(0.0,-)\end{array}\right]$&$\left[\begin{array}{c}(1.8,0.4)\\(54.2,9.7)\\(14.6,2.3)\\(23.0,1.8)\\(5.4,-)\\(0.3,-)\\(0.7,-)\\(0.0,-)\end{array}\right]$&$\left[\begin{array}{c}(0.0,0.0)\\(0.0,0.0)\\(0.0,0.0)\\(0.3,0.1)\\(0.1,0.0)\\(0.3,0.1)\\(8.7,-)\\(43.3,-)\end{array}\right]$&$\left[\begin{array}{c}(0.0,0.0)\\(0.0,-)\\(0.0,-)\\(0.2,-)\\(0.1,-)\\(0.3,-)\\(7.8,-)\\(38.8,-)\end{array}\right]$&$\left[\begin{array}{c}(37.7,7.5)\\(8.8,1.0)\\(14.1,1.1)\\(25.4,-)\\(4.1,-)\\(4.9,-)\\(3.6,-)\\(1.5,-)\end{array}\right]$&$\left[\begin{array}{c}(3.0,0.7)\\(0.0,0.0)\\(40.8,6.0)\\(20.2,0.5)\\(34.0,-)\\(0.2,-)\\(1.4,-)\\(0.4,-)\end{array}\right]$&$\left[\begin{array}{c}(0.2,0.1)\\(0.0,0.0)\\(0.8,0.2)\\(1.9,0.5)\\(1.5,0.4)\\(3.0,0.7)\\(64.9,9.0)\\(27.7,-)\end{array}\right]$&$\left[\begin{array}{c}26.6\\15.0\\14.9\\11.3\\7.8\\6.6\\9.0\\-\end{array}\right]$\\\hline
\multicolumn{15}{c}{$ccns\bar{s}$ states}\\\hline
&&$\Xi_c^*D_s^*$&$\Omega_{cc}^*K^*$&$\Xi_{cc}^*\phi$&&&&&&&&&&\\$\frac52^-$&$\left[\begin{array}{c}4697.0\\4642.6\end{array}\right]$&$\left[\begin{array}{c}(17.4,-)\\(15.9,-)\end{array}\right]$&$\left[\begin{array}{c}(96.4,7.0)\\(3.6,-)\end{array}\right]$&$\left[\begin{array}{c}(2.2,-)\\(97.8,-)\end{array}\right]$&&&&&&&&&&$\left[\begin{array}{c}7.0\\-\end{array}\right]$\\
&&$\Xi_c^*D_s^*$&$\Xi_c^*D_s$&$\Xi_c'D_s^*$&$\Xi_cD_s^*$&&$\Omega_{cc}^*K^*$&$\Omega_{cc}^*K$&$\Omega_{cc}K^*$&$\Xi_{cc}^*\phi$&$\Xi_{cc}^*\eta$&$\Xi_{cc}^*\eta'$&$\Xi_{cc}\phi$&\\
$\frac32^-$&$\left[\begin{array}{c}4713.5\\4669.8\\4619.8\\4604.6\\4553.2\\4441.9\\4274.0\end{array}\right]$&$\left[\begin{array}{c}(8.9,-)\\(4.3,-)\\(2.9,-)\\(29.2,-)\\(0.1,-)\\(0.8,-)\\(1.0,-)\end{array}\right]$&$\left[\begin{array}{c}(3.6,1.4)\\(0.6,0.2)\\(0.4,0.0)\\(15.2,-)\\(1.3,-)\\(17.9,-)\\(2.6,-)\end{array}\right]$&$\left[\begin{array}{c}(0.8,0.2)\\(3.6,-)\\(31.2,-)\\(1.4,-)\\(2.7,-)\\(4.2,-)\\(0.6,-)\end{array}\right]$&$\left[\begin{array}{c}(0.7,0.4)\\(2.2,0.8)\\(2.3,0.6)\\(0.3,0.0)\\(16.0,-)\\(4.7,-)\\(7.2,-)\end{array}\right]$&&$\left[\begin{array}{c}(64.5,6.1)\\(31.9,-)\\(1.2,-)\\(0.9,-)\\(1.5,-)\\(0.0,-)\\(0.0,-)\end{array}\right]$&$\left[\begin{array}{c}(0.0,0.0)\\(0.0,0.0)\\(0.0,0.0)\\(3.6,0.9)\\(0.4,0.1)\\(5.2,0.9)\\(90.7,1.0)\end{array}\right]$&$\left[\begin{array}{c}(9.0,1.4)\\(41.1,5.0)\\(17.1,1.0)\\(0.5,0.0)\\(31.4,-)\\(0.9,-)\\(0.0,-)\end{array}\right]$&$\left[\begin{array}{c}(15.5,0.7)\\(36.6,-)\\(22.5,-)\\(21.0,-)\\(3.1,-)\\(0.6,-)\\(0.7,-)\end{array}\right]$&$\left[\begin{array}{c}(0.3,0.1)\\(1.2,0.3)\\(0.8,0.2)\\(4.1,1.1)\\(2.0,0.5)\\(30.9,6.4)\\(7.9,0.7)\end{array}\right]$&$\left[\begin{array}{c}(0.4,0.0)\\(1.3,0.1)\\(0.9,-)\\(4.6,-)\\(2.3,-)\\(34.5,-)\\(8.8,-)\end{array}\right]$&$\left[\begin{array}{c}(15.7,2.1)\\(0.0,0.0)\\(17.1,-)\\(6.4,-)\\(52.3,-)\\(6.1,-)\\(2.3,-)\end{array}\right]$&$\left[\begin{array}{c}12.4\\6.4\\1.8\\2.0\\0.6\\7.3\\1.7\end{array}\right]$
\\
&&$\Xi_c^*D_s^*$&$\Xi_c'D_s^*$&$\Xi_{c}D_s^*$&$\Xi_c'D_s$&$\Xi_cD_s$&$\Omega_{cc}^*K^*$&$\Omega_{cc}K^*$&$\Omega_{cc}K$&$\Xi_{cc}^*\phi$&$\Xi_{cc}\phi$&$\Xi_{cc}\eta$&$\Xi_{cc}\eta'$&\\
$\frac12^-$&$\left[\begin{array}{c}4780.7\\4688.1\\4640.6\\4560.8\\4541.3\\4472.1\\4409.5\\4176.8\end{array}\right]$&$\left[\begin{array}{c}(21.4,3.8)\\(22.5,-)\\(3.6,-)\\(0.5,-)\\(5.8,-)\\(0.1,-)\\(1.5,-)\\(0.3,-)\end{array}\right]$&$\left[\begin{array}{c}(0.8,0.2)\\(18.0,-)\\(0.1,-)\\(0.3,-)\\(3.8,-)\\(1.1,-)\\(10.7,-)\\(1.3,-)\end{array}\right]$&$\left[\begin{array}{c}(0.2,0.0)\\(0.1,0.0)\\(6.6,2.0)\\(19,2,-)\\(0.5,-)\\(30.1,-)\\(1.4,-)\\(0.2,-)\end{array}\right]$&$\left[\begin{array}{c}(0.1,0.0)\\(0.2,0.0)\\(1.2,0.4)\\(3.3,0.4)\\(25.3,-)\\(4.8,-)\\(6.4,-)\\(0.4,-)\end{array}\right]$&$\left[\begin{array}{c}(0.0,0.0)\\(0.1,0.0)\\(0.7,0.4)\\(3.5,1.4)\\(1.2,0.4)\\(15.9,3.8)\\(9.0,-)\\(11.3,-)\end{array}\right]$&$\left[\begin{array}{c}(51.6,7.8)\\(3.3,0.2)\\(27.3,-)\\(6.4,-)\\(7.3,-)\\(4.0,-)\\(0.1,-)\\(0.0,-)\end{array}\right]$&$\left[\begin{array}{c}(0.4,0.1)\\(33.7,4.6)\\(29.9,2.7)\\(32.8,-)\\(1.4,-)\\(1.3,-)\\(0.6,-)\\(0.0,-)\end{array}\right]$&$\left[\begin{array}{c}(0.0,0.0)\\(0.1,0.0)\\(0.0,0.0)\\(0.4,0.1)\\(1.1,0.3)\\(0.4,0.1)\\(18.4,3.7)\\(79.5,-)\end{array}\right]$&$\left[\begin{array}{c}(34.7,4.6)\\(7.3,-)\\(20.8,-)\\(19.9,-)\\(3.4,-)\\(6.6,-)\\(6.6,-)\\(0.7,-)\end{array}\right]$&$\left[\begin{array}{c}(1.4,0.3)\\(7.0,0.8)\\(29.1,-)\\(18.4,-)\\(38.2,-)\\(1.4,-)\\(4.0,-)\\(0.4,-)\end{array}\right]$&$\left[\begin{array}{c}(0.1,0.0)\\(0.1,0.0)\\(0.9,0.3)\\(0.9,0.2)\\(4.2,1.2)\\(7.3,1.8)\\(21.6,4.9)\\(12.2,0.5)\end{array}\right]$&$\left[\begin{array}{c}(0.1,0.0)\\(0.1,0.0)\\(1.0,0.1)\\(1.0,-)\\(4.7,-)\\(8.2,-)\\(24.1,-)\\(13.6,-)\end{array}\right]$&$\left[\begin{array}{c}16.8\\5.6\\5.9\\2.1\\1.9\\5.7\\8.6\\0.5\end{array}\right]$\\\hline\hline\end{tabular}}\end{table}

\subsection{Spectra and decay widths of $ccns\bar{n}$ and $ccns\bar{s}$ states}

There are seventeen isovector $ccns\bar{n}$ pentaquark states. The number of their degenerate $I=0$ $ccns\bar{n}$ partners and that of $ccns\bar{s}$ states are also seventeen. The excited conventional doubly-charmed baryons may mix with these pentaquarks if the couplings are strong. Here, only the predicted pentaquark states are considered. We list the numerical results for them in tables \ref{massofccnsq} and \ref{decayofccnsq} and show their relative positions in Fig. \ref{figures} (g) and (h). From table \ref{decayofccnsq}, the decay properties of $I=1$ and $I=0$ $ccns\bar{n}$ states are different.

The $ccns\bar{n}$ pentaquarks are located from 3898.7 to 4759.1 MeV. Their masses are between the estimated values using the $\Xi_cD$ threshold and those using the $\Xi_{cc}\bar{K}$, and are closer to the latter values. The spin-parities of the highest and lowest states are both $\frac12^-$. The masses of $\frac52^-$ pentaquarks may be around 4589 MeV or 4665 MeV. In the $I=1$ case, the higher state has three main decay channels while the lower has two. The decay properties of the $I=0$ states are similar to those of $I=1$, except for replacing $\Omega_{cc}^*\rho$ with $\Omega_{cc}^*\omega$. For the $\Xi_c^*D^*$ and $\Xi_{cc}^*\bar{K}^*$ channels, the total isospin can be 1 and 0. The partial widths to one of the two channels for the degenerate isovector and isoscalar $ccns\bar{n}$ states are the same. Compared to the $\Omega_{cc}^{*}\rho$ channel for its degenerate $I=1$ partner, the partial width into the $\Omega_{cc}^{*}\omega$ channel for an $I=0$ pentaquark is almost the same. The reason is that the nearly mass-degenerate $\rho$ and $\omega$ have the same color-spin wavefunction, leading to equal phase spaces for the two channels.

The $\frac32^-$ $ccns\bar{n}$ states have seven energy levels which are between 3993.3 and 4696.4 MeV. In the $I=1$ case, the highest two states share similar dominant decay channels, except for replacing $\Xi_c^*D^*$ with $\Xi_c'D^*$ for the second highest state. The partial width ratios for these two states are
 \begin{eqnarray}
 	&&\Gamma_{\Xi_c^*D^*}:\Gamma_{\Xi_cD^*}:\Gamma_{\Xi_c^*D}:\Gamma_{\Omega_{cc}^*\rho}:\Gamma_{\Omega_{cc}\rho}:\Gamma_{\Xi_{cc}^*\bar{K}^*}:\Gamma_{\Xi_{cc}\bar{K}^*}:\Gamma_{\Xi_{cc}^*\bar{K}}\\\nonumber&=&20.0:4.0:20.0:99.0:30.0:28.0:39.0:1.0
 \end{eqnarray}
 and
 \begin{eqnarray}
 	&&\Gamma_{\Xi_cD^*}:\Gamma_{\Xi_c'D^*}:\Gamma_{\Xi_c^*D}:\Gamma_{\Omega_{cc}^*\rho}:\Gamma_{\Omega_{cc}\rho}:\Gamma_{\Xi_{cc}^*\bar{K}^*}:\Gamma_{\Xi_{cc}\bar{K}^*}:\Gamma_{\Xi_{cc}^*\bar{K}}\\\nonumber&=&6.0:4.0:1.0:25.0:37.0:24.5:1.5:1.0,
 \end{eqnarray}
respectively. The third highest state has five main channels, $\Xi_cD^*$, $\Xi_c^*D$, $\Omega_{cc}\rho$, $\Omega_{cc}^*\pi$, and $\Xi_{cc}\bar{K}^*$. The fourth highest state has three major channels, $\Xi_cD^*$, $\Omega_{cc}\rho$, and $\Xi_{cc}^*\bar{K}$. The main decay channels of the third lowest state are similar to them except for replacing $\Omega_{cc}\rho$ with $\Omega_{cc}^*\pi$. The second lowest state mainly decays into $\Omega_{cc}^*\pi$ and $\Xi_{cc}^*\bar{K}$ while the dominant decay channel for the lowest is $\Omega_{cc}^*\pi$. In the case of $I=0$, the possible decay channels are similar to those in the $I=1$ case, except for replacing $\Omega_{cc}^{(*)}\rho$ and $\Omega_{cc}^*\pi$ with $\Omega_{cc}^{(*)}\omega$ and $\Omega_{cc}^*\eta^{(')}$, respectively. If the isospin of a decay channel can be 1 and 0, one again gets the same partial width to that channel for the degenerate isovector and isoscalar $ccns\bar{n}$ states. The partial width into $\Omega_{cc}^{(*)}\rho$ for an $I=1$ pentaquark and that into $\Omega_{cc}^{(*)}\omega$ for its degenerate isoscalar state are almost the same. However, the situation is different for the pseudoscalar meson decays. The $\eta$ and $\eta'$ mesons are hundreds of MeV heavier than $\pi$. Their wavefunctions are also different. As a result, the partial widths into the $\Omega_{cc}^*\eta$ and $\Omega_{cc}^*\eta'$ channels for an $I=0$ pentaquark differ significantly from those into the $\Omega_{cc}^*\pi$ channel for its degenerate $I=1$ partner state. It is observed that the $\Omega_{cc}^*\eta$ and $\Omega_{cc}\eta'$ channels have small contributions to the rearrangement decays for the $I(J^P)=0(\frac32^-)$ states. Additionally, the ground $I(J^P)=0(\frac32^-)$ state should be very narrow.

The remaining eight energy levels belong to the $J^P=\frac12^-$ pentaquarks. One observes similar features to the $\frac32^-$ case. The possible decay channels of the $I=0$ states are the same as those of $I=1$ states except for replacing $\Omega_{cc}^{(*)}\rho$ and $\Omega_{cc}\pi$ with $\Omega_{cc}^{(*)}\omega$ and $\Omega_{cc}\eta^{(')}$, respectively. If two degenerate isoscalar and isovector pentaquarks have the same rearrangement decay channel, their partial widths in this channel are equal. When these two states have the  $\Omega_{cc}^{(*)}\omega$ channel or $\Omega_{cc}^{(*)}\rho$ channel, the corresponding partial widths are nearly equal. If an isoscalar $\frac12^-$ pentaquark can decay into $\Omega_{cc}\eta$ or $\Omega_{cc}\eta'$, this channel only has tiny contribution to its total width. All the $I=1$ and $I=0$, $J^P=\frac12^-$ states have strong decay channels except the lowest $I(J^P)=0(\frac12^-)$ pentaquark. This state should be stable if it really exists. In the study of Ref. \cite{Guo:2017vcf} within chiral effective theory, one $I(J^P)=0(\frac12^-)$ bound state about 150 MeV below the $\Xi_{cc}\bar{K}$ threshold was observed. Our stable pentaquark is about 210 MeV below this threshold.

Let us reconsider the $I_{nn}=1$ $ccnn\bar{n}$ case where the $I=\frac32$ and $\frac12$ states are also degenerate. The degenerate partner states have the same partial width for the same $(cnn)(c\bar{n})$-type decay channel while they have different coupling matrices and thus different partial widths for the same $(ccn)(n\bar{n})$-type channel. The reason is that the $I_{nn}=1$ portions in the $I=1$ and $I=0$ $(ccn)(n\bar{n})$ wave functions are different. The situation differs from the $ccns\bar{n}$ case.
  
The $ccns\bar{s}$ pentaquark states are located from 4176.8 to 4780.7 MeV. The spectrum is similar to the $ccns\bar{n}$ case. The higher $J^P=\frac52^-$ pentaquark has an S-wave decay channel $\Omega_{cc}^*K^*$, while the lower one may be narrow. The masses of the seven $J^P=\frac32^-$ states are  between 4274.0 MeV and 4713.5 MeV. The highest state has eight main decay channels, $\Xi_c^*D_s$, $\Xi_c'D_s^*$, $\Xi_cD_s^*$, $\Omega_{cc}^*K^*$, $\Omega_{cc}K^*$, $\Xi_{cc}^*\phi$, $\Xi_{cc}^*\eta$, and $\Xi_{cc}\phi$. The second highest state has five dominant channels, $\Xi_c^*D_s$, $\Xi_cD_s^*$, $\Omega_{cc}K^*$, $\Xi_{cc}^*\eta$, and $\Xi_{cc}^*\eta'$. The third highest mainly decays into $\Xi_cD_s^*$, $\Omega_{cc}K^*$, and $\Xi_{cc}^*\eta$. The $\Omega_{cc}^*K$ and $\Xi_{cc}^*\eta$ are two major decay modes for the lower four pentaquarks. The partial width ratios for these four states  from high-mass to low-mass are $\Gamma_{\Omega_{cc}^*K}:\Gamma_{\Xi_{cc}^*\eta}=0.82,\ 0.20,\ 0.14,\ \text{and} \ 1.43$, respectively. There are eight $J^P=\frac12^-$ pentaquarks. All possible rearrangement decay channels are open for the highest state, but only six are dominant, $\Xi_c^*D_s^*$, $\Xi_c'D_s^*$, $\Omega_{cc}^*K^*$, $\Omega_{cc}K^*$, $\Xi_{cc}^*\phi$, and $\Xi_{cc}\phi$. The second highest state features three major channels $\Omega_{cc}^*K^*$, $\Omega_{cc}K^*$, and $\Xi_{cc}\phi$. The third highest has six dominant channels, $\Xi_cD_s^*$, $\Xi_c'D_s$, $\Xi_cD_s$, $\Omega_{cc}K^*$, $\Xi_{cc}\eta$, and $\Xi_{cc}\eta'$. The fourth highest state can decay into $\Xi_c'D_s$, $\Xi_cD_s$, $\Omega_{cc}K$, and $\Xi_{cc}\eta$. The next two states have masses 4541.3 MeV and 4472.1 MeV, respectively. They share the same dominant decay channels, with partial width ratios $\Gamma_{\Xi_cD_s}:\Gamma_{\Omega_{cc}K}:\Gamma_{\Xi_{cc}\eta}=1.3:1.0:4.0$ and $38.0:1.0:18.0$, respectively. The second lowest state has two dominant channels $\Omega_{cc}K$ and $\Xi_{cc}\eta$, while the lowest state which is slightly above the $\Xi_{cc}\eta$ threshold mainly decays into $\Xi_{cc}\eta$. The hadron-level study of Ref. \cite{Guo:2017vcf} also suggests the existence of a resonance peak close to $\Xi_{cc}\eta$ in the  (strangeness=0, isospin=1/2) channel.

%==================================================
\section{Discussions and summary}\label{secIV}
%==================================================

In this work, we investigate the spectra of S-wave doubly-charmed pentaquark states in the compact $ccqq\bar{q}$ configuration by employing a simple CMI model. We assume that the $P_\psi^N(4312)^+$ is a compact hidden-charm pentaquark with $I(J^P)=\frac12(\frac32^-)$ and treat it as a reference to estimate masses of the $ccqq\bar{q}$ states. This assumption is from the consistency requirement for an interpretation of the observed hidden-charm pentaquark states \cite{Cheng:2019obk,Li:2023aui}. Compared with Ref. \cite{Zhou:2018bkn} in which the doubly-charmed pentaquark masses estimated with the reference (charmed baryon)-(charmed meson) thresholds were adopted as the predicted values, the results given here are tens of MeV lower in most cases. The $ccss\bar{s}$ masses are about 150 MeV lower. 

In the CMI model without any dynamics, the numerical calculation of multiquark masses relies on the values of effective quark masses $m_i$'s and coupling constants $C_{ij}$'s. They are extracted from conventional mesons and baryons. However, they may introduce uncertainties into our results. The reason is that the conventional hadrons and compact pentaquarks have different inner structures. There is the possibility that the obtained effective quark masses and coupling parameters may not be directly suitable for the multiquark systems. The results of Refs. \cite{Liu:2019zoy,Zhou:2018bkn,Li:2023wxm,Cheng:2020nho,Wu:2018xdi} have illustrated this problem. Since the values of effective quark masses are much larger than those of coupling constants and the former induce more uncertainties, one can partially reduce the model uncertainty by modifying the original mass formula. Considering the compact configuration and quark contents, we introduce the reference pentaquark $P_\psi^N(4312)^+$ into the formula. In principle, the effects of coupling parameters should not be ignored, either. At present, we do not consider such effects because of the lack of adequate awareness for multiquark hadrons in extracting parameters.

In the adopted mass formula, we have $\tilde{m}_{penta}=4382.6$ MeV for the pentaquark mass estimation while $\tilde{m}_{tetra}=4231.1$ MeV for the tetraquark mass estimation \cite{Li:2023wug}. They play the role of $\sum_i m_i$ in the original mass formula but with a modified value in which the missed effect is partially compensated. We take a look at the consistency problem between the tetraquark and pentaquark studies. The difference in quark numbers is one between these two cases. If one takes $\tilde{m}_{penta}=2m_c+3m_n$ and $\tilde{m}_{tetra}=2m_c+2m_s$, we get $\tilde{m}_{penta}-\tilde{m}_{tetra}=m_n-2\Delta_{sn}$ and then
$m_n=\tilde{m}_{penta}-\tilde{m}_{tetra}+2\Delta_{sn}=332.7$ MeV. This is an acceptable value for the effective mass of light constituent quark which includes uncertainty corrections. %Note that it had been suspected that ....Stancu??

According to our estimations, there are three stable doubly-charmed pentaquark states, the lowest $I(J^P)=\frac12(\frac12^-)$ $ccnn\bar{n}$ with $I_{nn}=0$, the lowest $0(\frac12^-)$ $ccnn\bar{s}$, and the lowest $0(\frac12^-)$ $ccns\bar{n}$. Their internal charm quarks can be a color $6$ or $\bar{3}$ state, as shown in Eq. \eqref{color}, and both representations contribute to a compact multiquark. From the obtained wave functions listed in tables \ref{massofccnnn}, \ref{massofccnns}, and \ref{massofccnsq}, one finds that the color $\bar{3}$ component contributes dominantly to these three states, with the proportions being 0.99, 0.97, and 0.99, respectively. In terms of spin, the $qq\bar{q}$ cluster can be a spin $\frac32$ or $\frac12$ state. We find that $s_{qq\bar{q}}=\frac12$ is the dominant component for these stable states, with the propotions being 0.69, 0.71, and 0.68, respectively. Just from color and spin representations, they look like analogies of $\Xi_{cc}$ by replacing the $n$ quark with the triquark $qq\bar{q}$.

Besides the spectrum, we also consider the dominant two-body strong decays of doubly-charmed pentaquark states in a straightforward rearrangement model. We assume that the Hamiltonian governing the decay process is a constant which is extracted from the measured width of $P_\psi^N(4312)^+$. Utilizing this extracted parameter, we calculate the partial widths for corresponding decay channels of the studied $ccqq\bar{q}$ states. Since different states have distinct constituent quark components and inner interactions, the adopted assumption is certainly very crude and the differences for decay parameters could not be overlooked. Consequently, the  estimated values of partial widths may deviate from experimental measurements and are intended solely for a reference. Nevertheless, the partial width ratios for different channels of a specific state are more indicative. Once a doubly-charmed pentaquark-like state could be observed, the quantum number, mass, and width information for the predicted states may be used to understand its structure. 
% By considering quantum numbers, locations of spectrum, and ratios of partial decay widths, one can identify possible compact doubly-charmed pentaquarks from experiments.
	
To summarize, we studied the spectra and decay properties of the compact doubly-charmed pentaquark states. Their masses were estimated in a modified CMI model whose mass formula includes the scale related to the reference pentaquark state $P_\psi^N(4312)^+$ with the assumed $I(J^P)=\frac12(\frac32^-)$. We also obtained estimations of their two-body strong decay widths and related partial width ratios in a simple rearrangement scheme. Three stable states were found in the present work, the lowest $I(J^P)=\frac12(\frac12^-)$ $ccnn\bar{n}$ with $I_{nn}=0$, the lowest isoscalar $J^P=\frac12$ $ccnn\bar{s}$, and the lowest $J^P=\frac12^-$ $ccns\bar{n}$ with $I=0$. We hope that our predictions can help future investigations on properties of the doubly-charmed pentaquark states.

%Our predictions may be used to understand whether an observed state has dominant compact pentaquark components.

%*********************************************
\section{ACKNOWLEDGMENTS}
%*********************************************

This project was supported by the National Natural Science Foundation of China (Nos. 12235008, 12275157, 12475143) and the Shandong Province Natural Science Foundation (ZR2023MA041).\\

{\bf Data Availability Statement} No Data were associated with the manuscript.
%\input{latex(EU).tex}
%\end{document}
%	\bibliographystyle{unsrt}
%	\bibliography{references}

\end{document}